\documentclass{amsart}
\usepackage{amssymb}

\begin{document}

\title[Energy inequalities]{Energy inequalities for a model of wave propagation in cold plasma}
\author{Thomas H. Otway}
\address{Department of Mathematics,
Yeshiva University, 500 W 185th Street, New York, New York 10033}
\email{otway@yu.edu}
\date{}
\begin{abstract}
Energy inequalities are derived for an elliptic-hyperbolic operator
arising in plasma physics. These inequalities imply the existence of
distribution and weak solutions to various closed boundary-value
problems. An existence theorem is proven for a related class of
Keldysh equations, and the failure of expected methods for obtaining
uniqueness is discussed. The proofs use ideas recently introduced by
Lupo, Morawetz, and Payne for a generalized Tricomi operator. The
existence of strong solutions under open boundary conditions is also
proven. \textit{MSC2000}: 35M10, 35D05, 82D10.

\medskip

\end{abstract}
\maketitle

\section{Introduction}

The equation

\begin{equation}\label{cold}
    \left( x-y^2 \right)u_{xx} + u_{yy}=0
\end{equation}
arises in models of wave propagation through a linear dielectric
medium (``cold plasma") at frequencies lying below the geometrical
optics range; for the physical context, see \cite{W2}. Here
$u(x,y),$ $\left(x,y\right)\in\mathbb{R}^2,$ is a scalar function.
A subscripted variable denotes partial differentiation in the
direction of the variable.

The significant property of eq.\ (\ref{cold}) is that it changes
from elliptic to hyperbolic type along the parabola

\[
x-y^2=0.
\]
By analogy with the equations of steady flow, which change from
elliptic to hyperbolic type at the speed of sound, it has become
conventional to call this parabola the \emph{sonic} curve; in the
context of the cold plasma model it is also called a
\emph{resonance} curve. Except for a point at the origin, eq.\
(\ref{cold}) can be mapped into an equation having the same general
form as the Tricomi equation

\begin{equation}\label{tricomi}
    yu_{xx} + u_{yy}=0,
\end{equation}
an equation which is somewhat more accessible than (\ref{cold}).
However, both the physical and mathematical interest of eq.\
(\ref{cold}) arise from the tangency of the sonic curve to the
line $x=0$ at the origin. This is the point at which plasma
heating might occur in the physical model, and a point which
appears to be singular in numerical studies of solutions; see
\cite{MSW}, \cite{PF}, and \cite{W1}.

A variety of lower-order terms have been affixed to eq.\
(\ref{cold}) in the literature; see, \emph{e.g.}, \cite{MSW},
\cite{PF}, and \cite{Y}. The variants reflect different toy models
for the equation satisfied by the field potential; compare
equations (2) and (9) of \cite{PF}. The variants tend to have the
general form

\begin{equation}\label{formal}
    \left(x-y^2\right)u_{xx}+u_{yy}+\kappa u_x=0,
\end{equation}
where the constant $\kappa$ lies in a specified interval. Note
that if $\kappa=1,$ then the associated differential operator is
formally self-adjoint.

The formulation of boundary-value problems for eq.\ (\ref{formal})
is of considerable interest, as the boundary conditions which are
physically natural do not appear to be mathematically natural. In
particular, it is shown in \cite{MSW} that the \emph{closed}
Dirichlet problem, in which the solution is prescribed on the
entire boundary, is over-determined for $C^2$ solutions of
(\ref{formal}) with $\kappa=1/2.$ However, the physical properties
of electromagnetic waves in the cold plasma model suggest that
closed boundary-value problems are natural and should be correctly
posed for eq.\ (\ref{formal}).

In Secs.\ 2 and 3 we prove that solutions to closed boundary-value
problems for (\ref{formal}) do in fact exist. Although we do not
expect classical solutions, we show in Sec.\ 2 the existence of
distribution solutions to a homogeneous Dirichlet problem, under
minimal hypotheses on the domain boundary. In fact these solutions
are somewhat smoother than conventional distribution solutions, as
they lie in the function space $L^2.$ An even smoother
distribution solution, which lies in a weighted Sobolev space, is
derived in Sec.\ 4 under stronger hypotheses on the boundary. We
prove the existence of conventional weak solutions to a class of
closed boundary-value problems, under strong hypotheses on the
boundary, in Sec.\ 3.

In Sec.\ 4 we also consider a simpler case in which eq.\
(\ref{cold}) is replaced by a generalized Cinquini-Cibrario
equation \cite{C}

\begin{equation}\label{cibrario}
    x^{2k+1}u_{xx} + u_{yy}+\mbox{lower order}=0,
\end{equation}
where $k$ is a non-negative integer. (In fact we do not treat the
case $k=0,$ which is the case initially studied by
Cinquini-Cibrario.) Equation (\ref{cibrario}) has independent
mathematical interest as one of two classes of normal forms to
which linear, second-order elliptic-hyperbolic equations can be
reduced near a point \cite{Bi}. In terms of the physical model,
eq.\ (\ref{cibrario}) would correspond roughly to a resonance
surface which coincides with a flux surface. In this case the
plasma behaves like a perpendicular stratified medium and energy
absorption occurs along the entire surface, a situation more
amenable to standard physical arguments than the case in which the
two surfaces are tangent at a single point; see p. 42 of \cite{W1}
for a discussion.

Equation (\ref{cibrario}) is an example of an equation of
so-called \emph{Keldysh type} $-$ that is, an equation of the form

\begin{equation}\label{keldysh}
    K(x)u_{xx}+u_{yy}+\mbox{lower order}=0,
\end{equation}
where $K(x)$ is a continuously differentiable function such that
$K(0)=0$ and $xK(x) > 0$ for $x \neq 0.$ Equations of this kind,
with various lower-order terms, arise in transonic fluid dynamics
and singular optics (see, \emph{e.g.,} Sec.\ 3 of \cite{CK}, and
\cite{MT}). The arguments of Sec.\ 4 apply, with only notational
alterations, to the slightly more general case of a type-change
operator having the form $K(x)=x|x|^{n-1}$ for $n\in \mathbb{R},$
$n>3.$ (The proofs require $n$ to be large enough for $K$ to have
three continuous derivatives.)

In Sec.\ 5 we obtain the existence of unique solutions to eq.\
(\ref{formal}), but under \emph{open} boundary conditions, in
which the solution is prescribed on a proper subset of the
boundary. In Sec.\ 5.1 we briefly discuss the failure of the
preceding methods to provide uniqueness in an obvious way.
Conditions for the existence of a unique, strong solution are
introduced in Sec.\ 5.2. The existence of weak $L^2$ solutions to
a large class of open boundary-value problems is proven in Sec.\ 6
by extending the arguments of \cite{O}.

The approach taken in Secs.\ 2-4 is modeled on recent work by
Lupo, Morawetz, and Payne \cite{LMP}, \cite{P2} on the class of
equations

\begin{equation}\label{t-type}
    K(y)u_{xx}+u_{yy}=0,
\end{equation}
where $K(y)$ is a continuously differentiable function for which
$K(0)=0$ and $yK(y) > 0$ for $y \neq 0.$ In the special case
$K(y)=y,$ (\ref{t-type}) reduces to the Tricomi equation. For this
reason, equations of the form (\ref{t-type}) are said to be of
\emph{Tricomi type.}

We adapt the ideas of \cite{LMP} to the cold plasma context by
proving inequalities for (\ref{formal}) having the form

\begin{equation}\label{energy}
    ||L^\ast v||_U \geq C||v||_V,
\end{equation}
where $C$ is a positive constant, $U$ and $V$ are function spaces,
and $L$ is the differential operator of (\ref{formal}) with
adjoint $L^\ast.$ These \emph{energy inequalities} are used to
show the existence of a solution to boundary-value problems in an
appropriate function space (see, \emph{e.g.}, \cite{B}, Ch.\ 2).
However, we state the inequalities as theorems, rather than as
lemmas, and derive the existence theorems as corollaries. We do
this because in their exploitation of inequality (\ref{energy}),
the arguments for eqs.\ (\ref{formal}) and (\ref{t-type}) are
essentially the same. (While differential operators of Tricomi
type are formally self-adjoint, the extension of the existence
arguments in \cite{LMP} to the non-self-adjoint case is standard;
see, \emph{e.g.}, the existence proof in \cite{O} for weak
solutions to a Dirichlet problem for (\ref{formal}) under open
boundary conditions.) The arguments for the two equations differ,
however, in their derivations of the energy inequality itself,
which depend on the form of the type-change function $K.$ One of
the main problems of this paper is to find multipliers which allow
the Friedrichs $abc$ method to be applied in the right way, either
in its original form or in the more recent integral variant
introduced by Didenko \cite{D}. Another is to establish \emph{a
priori} restrictions on the domain which allow the method of
energy inequalities to be applied. (In Sec.\ 5.2 we adopt a third
approach, also due to Friedrichs, in order to establish sufficient
conditions for uniqueness.)

In addition to its physical interest, the existence of solutions
to closed boundary-value problems for equations of the form
(\ref{formal}) and (\ref{cibrario}) has purely mathematical
interest as an extension of the methods introduced in \cite{LMP}
to equations which are not of Tricomi type. The existence of
solutions to closed boundary-value problems for certain
elliptic-hyperbolic systems which are not of Tricomi type is shown
in \cite{G} and \cite{MT} under special conditions on the boundary
that do not require the methods of \cite{LMP}. Our results suggest
that the existence $-$ but not necessarily uniqueness $-$ of
distribution and weak solutions for equations of Keldysh type can
be shown by arguments closely modeled on those of \cite{LMP}.

The following hypotheses on the domain $\Omega$ are assumed
throughout: It is a bounded, connected domain of $\mathbb{R}^2$
having piecewise smooth boundary $\partial\Omega,$ oriented in a
counterclockwise direction; the domain includes both an arc of the
sonic curve and the origin of coordinates in $\mathbb{R}^2$ (so
that, for example, eq.\ (\ref{formal}) is elliptic-hyperbolic but
not equivalent to an equation of Tricomi type). We neither require
nor exclude the existence of characteristic arcs on the boundary.
We use the term \emph{elliptic boundary} to refer to points
$\left(x,y\right)$ of the domain boundary on which the type-change
function $K\left(x,y\right)$ is positive. Similarly, by the
\emph{hyperbolic boundary} we mean boundary points for which the
type-change function is negative.

\section{Inequalities leading to distribution solutions}

The function spaces introduced in \cite{D} and \cite{LMP} reappear
in this paper with $|K(y)|$ replaced by a different weight
function, also denoted by $K.$ In particular, we define the space
$H^1_0(\Omega; K)$ to be the closure of $C_0^\infty(\Omega)$ with
respect to the norm

\[
||u||_{H^1(\Omega; K)}=\left[\int\int_{\Omega}
\left(|K|u_x^2+u_y^2+u^2\right)\,dxdy\right]^{1/2},
\]
where $|K|=\left|x-y^2\right|.$ We can write the $H^1_0(\Omega;
K)$-norm in the form

\[
||u||_{H^1_0(\Omega; K)}=\left[\int\int_\Omega
\left(|K|u_x^2+u_y^2\right)\,dxdy\right]^{1/2}
\]
as a consequence of the weighted Poincar\'{e} inequality

\begin{equation}\label{poin}
    ||u||^2_{L^2(\Omega)} \leq C\int\int_\Omega
\left(|K|u_x^2+u_y^2\right)\,dxdy.
\end{equation}
Here and below we denote by $C$ generic positive constants, the
value of which may change from line to line.

The complexity of the existence arguments is not increased if we
replace (\ref{formal}) by the inhomogeneous equation

\begin{equation}\label{f-alt1}
    Lu=f,
\end{equation}
where $f$ is a given, sufficiently smooth function of $(x,y)$ and

\begin{equation}\label{f-alt2}
    L=\left(x-y^2\right)\frac {\partial^2} {\partial x^2}+\frac{\partial^2}{\partial y^2}+ \kappa\frac{\partial}{\partial x}.
\end{equation}
By a \emph{distribution solution} of equations (\ref{f-alt1}),
(\ref{f-alt2}) with the boundary condition

\begin{equation}\label{boundary}
    u(x,y)=0\,\forall (x,y)\in\partial\Omega
\end{equation}
we mean a function $u\in L^2(\Omega)$ such that $\forall \xi \in
H^1_0(\Omega;K)$ for which $L^\ast\xi\in L^2(\Omega),$ we have

\begin{equation}\label{ds}
    \left(u,L^\ast\xi\right)=\langle f,\xi \rangle.
\end{equation}
Here $(\,,\,)$ denotes the $L^2$ inner product on $\Omega$ and
$\langle \,,\, \rangle$ is the \emph{duality bracket} associated
to the $H^{-1}$ norm \cite{L}

\[
||w||_{H^{-1}(\Omega;K)}=\sup_{0\neq\xi\in
C^\infty_0(\Omega)}\frac{|\langle w,\xi
\rangle|}{||\xi||_{H^1_0(\Omega;K)}}.
\]
Such a solution is a little smoother than the usual notion of
distribution solution, in which the solution fails to lie in a
true function space.

\bigskip

\textbf{Theorem 1}. \emph{Every $u\in C^2_0(\Omega)$ satisfies the
inequality}

\[
||u||_{H^1_0(\Omega;K)}\leq C||Lu||_{L^2(\Omega)},
\]
\emph{where $L$ is defined by (\ref{f-alt2}) with $\kappa \in
\left[0,2\right],$ and $K=x-y^2.$}

\bigskip

\emph{Proof}. We consider two cases.

\emph{Case 1:} $1\leq\kappa\leq 2.$ Let $\delta$ be a small,
positive constant. Define an operator $M$ by the identity

\[
    Mu=au+bu_x+cu_y
\]
for $a=-1,$ $c=2\left(2\delta-1\right)y,$ and

\[
b= \left\{
        \begin{array}{cr}
    \exp\left(2\delta K/Q_1\right) & \mbox{if $\left(x,y\right)\in\Omega^+$} \\

    \exp\left(6\delta K/Q_2\right) & \mbox{if $\left(x,y\right)\in\Omega^-$}\\
    \end{array}
    \right.,
\]
where

\[
\Omega^+=\left\lbrace\left(x,y\right)\in\Omega\,|\,K >
0\right\rbrace
\]
and $\Omega^-=\Omega\backslash\Omega^+.$ Choose
$Q_1=\exp\left(2\delta\mu_1\right),$ where

\[
\mu_1=\max_{\left(x,y\right)\in\overline{\Omega^+}}K.
\]
Then $\forall\left(x,y\right)\in\Omega^+,$ we have

\[
2\delta K\leq2\delta\mu_1\leq 2\delta\mu_1e^{2\delta\mu_1}=Q_1\log
Q_1.
\]
Dividing by $Q_1$ and exponentiating both sides, we conclude that
$b\leq Q_1$ on $\Omega^+.$ Define the negative number $\mu_2$ by

\[
\mu_2=\min_{\left(x,y\right)\in\overline{\Omega^-}}K
\]
and let $Q_2=\exp\left(\mu_2\right).$ Then $0<Q_2<1$ and, for
given $\Omega,$ we can choose $\delta$ to be so small that
$6\delta<Q_2.$ In that case, $\forall\left(x,y\right)\in\Omega^-,$

\[
6\delta K\geq 6\delta\mu_2 = 6\delta\log Q_2> Q_2\log Q_2.
\]
We conclude that $b> Q_2$ on $\Omega^-.$

We will estimate the quantity $(Mu,Lu)$ from above and below.
Integrating by parts, we have

\[
(Mu,Lu)=\int\int_\Omega \sum_{i=1}^9 \tau_i\,dxdy,
\]
where

\[
\tau_1=\left(auKu_x\right)_x-a_xKuu_x-\frac{1}{2}\left(au^2\right)_x+\frac{a_x}{2}u^2-aKu_x^2;
\]
\[
\tau_2=\left(auu_y\right)_y-a_yuu_y-au_y^2;
\]
\[
\tau_3=\frac{1}{2}\left(bKu_x^2\right)_x
-\frac{1}{2}\left(b_xK+b\right)u_x^2;
\]
\[
\tau_4=\left(bu_xu_y\right)_y-b_yu_xu_y-\frac{1}{2}\left(bu_y^2\right)_x+\frac{b_x}{2}u_y^2;
\]
\[
\tau_5=\left(cu_yKu_x\right)_x-c_xKu_xu_y
\]
\[
-\frac{1}{2}\left(cKu_x^2\right)_y-cu_xu_y+\left(\frac{c_y}{2}K-cy\right)u_x^2;
\]
\[
\tau_6=\frac{1}{2}\left[\left(cu_y^2\right)_y-c_y u_y^2\right];
\]
the lower-order terms are:
\[
\tau_7=\frac{\kappa}{2}\left[\left(au^2\right)_x-\kappa a_x
u^2\right];
\]
\[
\tau_8= \kappa bu_x^2;
\]
\[
\tau_9= \kappa c u_yu_x.
\]

As in the Tricomi case considered in \cite{LMP}, one of the
coefficients in $Mu$ fails to be continuously differentiable on
all of $\Omega.$ When integrating this quantity, a cut should be
introduced along the line $K=0$ separating $\Omega^+$ from
$\Omega^-.$ The boundary integrals involving $a,$ $b,$ and $c$ on
either side of this line will cancel by continuity.

We find that the boundary terms vanish by the compact support of
$u,$ and

\[
\left(Mu,Lu\right)=\int\int_{\Omega^+\cup\Omega^-}\omega
u^2+\alpha u_x^2+2\beta u_xu_y+\gamma u_y^2\,dxdy,
\]
where $\omega=0;$

\[
\alpha =
\left(\frac{c_y}{2}-a-\frac{b_x}{2}\right)K+\left(\kappa-\frac{1}{2}\right)b-cy,
\]
for

\[
\alpha_{|\Omega^+}=\left(2 -\frac{b}{Q_1}\right)\delta
K+2\left(1-2\delta\right)y^2+\left(\kappa-\frac{1}{2}\right)b
\]
and

\[
\alpha_{|\Omega^-} =
\left(3\frac{b}{Q_2}-2\right)\delta|K|+2\left(1-2\delta\right)y^2+\left(\kappa-\frac{1}{2}\right)b;
\]

\[
\beta=\frac{1}{2}\left[c\left(\kappa-1\right)-b_y\right] = \left\{
        \begin{array}{cr}
    y\left[2\delta \left(b/Q_1\right)+\left(\kappa-1\right)\left(2\delta-1\right)\right]\leq |y| & \mbox{in $\Omega^+$} \\

    y\left[6\delta \left(b/Q_2\right)+\left(\kappa-1\right)\left(2\delta-1\right)\right]\leq \kappa|y| & \mbox{in $\Omega^-$}\\
    \end{array}
    \right.;
\]

\[
\gamma=\frac{1}{2}\left(b_x-c_y\right)-a =\left\{
\begin{array}{cr}
    2\left(1-\delta\right)+\delta\left(b/Q_1\right) & \mbox{in $\Omega^+$} \\

    2\left(1-\delta\right)+3\delta\left(b/Q_2\right) & \mbox{in $\Omega^-$}\\
    \end{array}
    \right..
\]
On $\Omega^+,$ for any scalars $\xi$ and $\eta,$ we have by
Cauchy's inequality

\[
2\beta\xi\eta\leq y^2\xi^2+\eta^2,
\]
so

\[
\alpha\xi^2+2\beta\xi\eta+\gamma\eta^2\geq
\alpha\xi^2-\left(y^2\xi^2+\eta^2\right)+\gamma\eta^2=
\]

\[
\left[\left(2-\frac{b}{Q_1}\right)\delta
K+\left(1-4\delta\right)y^2+\left(\kappa-\frac{1}{2}\right)b\right]\xi^2+\left[\left(1-2\delta\right)+\frac{6b}{Q_1}\right]\eta^2
\]

\[
\geq \delta\left(K\xi^2+\eta^2\right),
\]
provided $\delta$ is sufficiently small. On $\Omega^-,$

\[
2\beta\xi\eta \leq 2\left|\kappa y\xi\eta\right|\leq
2\left(y^2\xi^2+\eta^2\right),
\]
so

\[
\alpha\xi^2+2\beta\xi\eta+\gamma\eta^2\geq
\alpha^2\xi^2-2\left(y^2\xi^2+\eta^2\right)+\gamma\eta^2=
\]

\[
\left[\left(3\frac{b}{Q_2}-2\right)\delta|K|-4\delta
y^2+\left(\kappa -
\frac{1}{2}\right)b\right]\xi^2+\delta\left(3\frac{b}{Q_2}-2\right)\eta^2
\]

\[
\geq \delta\left(|K|\xi^2+\eta^2\right).
\]
In estimating the coefficient of $\eta^2,$ we used the fact that
$b>Q_2$ on $\Omega^-.$ In estimating the coefficient of $\xi^2,$
we used the fact that

\[
\left(\kappa-\frac{1}{2}\right)b-4\delta y^2\geq
\frac{b}{2}-4\delta y^2\geq \frac{1}{2}e^{\mu_2}-4\delta y^2,
\]
which exceeds zero for $\delta$ sufficiently small with respect to
$\exp\left(-|\mu_2|\right)$ and $\max_{\overline\Omega} y^2.$

Integrating over each subdomain, we obtain

\begin{equation}\label{lower-est}
    \left(Mu,Lu\right)\geq \delta\int\int_\Omega\left(
|K|\,u_x^2+u_y^2\right)dxdy=\delta ||u||^2_{H^1_0(\Omega;K)}.
\end{equation}

We want to obtain an upper bound for $(Mu,Lu)$ in terms of the
$L^2$-norm of $Lu.$ First we recall that
$\max_{\left(x,y\right)\in\Omega}b\leq Q_1$ and estimate

\[
    (Mu,Lu)\leq \int\int_{\Omega}|u||Lu|\,dxdy+
\]

\begin{equation}\label{upper-est}
C\int\int_{\Omega}\left(Q_1|u_x|+|y||u_y|\right)|Lu|\,dxdy=
i_1+i_2.
\end{equation}
The Schwarz inequality and inequality (\ref{poin}) imply that

\[
i_1 \leq ||u||_{L^2(\Omega)}||Lu||_{L^2(\Omega)} \leq
C||u||_{H^1_0(\Omega;K)}||Lu||_{L^2(\Omega)}.
\]
Similarly,

\[
i_2\leq
C\left[\int\int_{\Omega}\left(|u_x|+|u_y|\right)^2\,dxdy\right]^{1/2}||Lu||_{L^2(\Omega)}
\]

\[
\leq C'||u||_{H^1_0(\Omega)}||Lu||_{L^2(\Omega)}.
\]

We claim that we can choose $\delta$ to be so small that
$b/2\geq\delta$ on $\Omega.$ On $\Omega^+$ this is obvious, as
$b\geq 1$ there. On $\Omega^-$ it is almost as obvious, as
$b>Q_2>6\delta>2\delta.$ Thus we have the additional estimate

\[
\alpha \geq \delta|K|+\frac{b}{2}\geq \delta,
\]
implying that

\begin{equation}\label{additional}
    \left(Mu,Lu\right)\geq \delta ||u||^2_{H^1_0(\Omega)}.
\end{equation}
Substituting the estimates for $i_1$ and $i_2$ into inequality
(\ref{upper-est}) and combining this with (\ref{additional}), we
obtain

\[
\delta||u||^2_{H^1_0(\Omega)}\leq \left(Mu,Lu\right)\leq
\]

\begin{equation}\label{H1bound}
C\left(||u||_{H^1_0(\Omega;K)}+||u||_{H^1_0(\Omega)}\right)||Lu||_{L^2(\Omega)}\leq
C'||u||_{H^1_0(\Omega)}||Lu||_{L^2(\Omega)}.
\end{equation}
Dividing eq.\ (\ref{H1bound}) by the $H^1_0$-norm of $u,$ we find
that

\[
||u||_{H^1_0(\Omega)}\leq C||Lu||_{L^2(\Omega)}.
\]
This completes the proof for the case $\kappa\in\left[1,2\right],$
as the norm on the left can be replaced by the corresponding
weighted norm. (Alternatively, substitute this estimate into the
extreme right-hand side of (\ref{H1bound}) and replace the extreme
left-hand side of (\ref{H1bound}) by the extreme right-hand side
of (\ref{lower-est}), to obtain

\[
||u||^2_{H^1_0(\Omega;K)}\leq C||Lu||_{L^2(\Omega)}^2;
\]
then take the square root of each side.)

\bigskip

\emph{Case 2:} $0\leq\kappa <1.$ Again subdivide the domain into
$\Omega^+$ and $\Omega^-$ by introducing a cut along the curve
$K=0.$ Integrate by parts as in case 1, choosing $a=-1;$

\[
b=\left\{
\begin{array}{cr}
    -NK & \mbox{in $\Omega^+$} \\

    NK & \mbox{in $\Omega^-$}\\
    \end{array}
    \right.,
\]
where $N$ is a constant satisfying

\begin{equation}\label{range}
    \frac{1+\tilde\delta}{3-\kappa}<N<\frac{1-\tilde\delta}{\kappa+1}
\end{equation}
for a sufficiently small positive constant $\tilde\delta,$ and

\[
c=-4Ny.
\]
The boundary integrals involving $a$ and $c$ on either side of the
curve will cancel and the boundary integrals involving $b$ will be
zero on the curve.

On $\Omega^+,$

\[
\alpha = \left[1-\left(1+\kappa\right)N\right]K+4Ny^2;
\]

\[
\beta = N\left(1-2\kappa\right)y;
\]

\[
\gamma=1+\frac{3}{2}N.
\]
Also,

\[
2\beta\xi\eta\geq -2N\left|y\xi\eta\right|\geq
-N\left(y^2\xi^2+\eta^2\right)
\]
as $\kappa\in[0,1),$ so

\[
\alpha\xi^2+2\beta\xi\eta+\gamma\eta^2\geq
\alpha\xi^2-N\left(y^2\xi^2+\eta^2\right)+\gamma\eta^2=
\]

\[
\left\lbrace\left[1-\left(1+\kappa\right)N\right]K+3Ny^2\right\rbrace\xi^2+\left(1+\frac{N}{2}\right)\eta^2\geq
\tilde\delta\left(K\xi^2+\eta^2\right),
\]
by condition (\ref{range}).

On $\Omega^-,$

\[
\alpha =-\left[1-\left(3-\kappa\right)N\right]|K|+4Ny^2;
\]

\[
\beta = N\left(3-2\kappa\right)y;
\]

\[
\gamma=1+\frac{5}{2}N.
\]
Because

\[
2\beta\xi\eta\geq -6N\left|y\xi\eta\right|,
\]
we have

\[
\alpha\xi^2+2\beta\xi\eta+\gamma\eta^2\geq
\]

\[
\left\lbrace-\left[1-\left(3-\kappa\right)N\right]|K|+4Ny^2\right\rbrace\xi^2-3N\left(y^2\xi^2+\eta^2\right)+\left(1+\frac{5}{2}N\right)\eta^2=
\]

\[
\left\lbrace-\left[1-\left(3-\kappa\right)N\right]|K|+Ny^2\right\rbrace\xi^2+\left(1-\frac{N}{2}\right)\eta^2\geq
\tilde\delta\left(|K|\xi^2+\eta^2\right),
\]
again using (\ref{range}).

We conclude that

\begin{equation}\label{new-low-est}
    (Mu,Lu) \geq \tilde\delta||u||^2_{H^1_0(\Omega;K)}.
\end{equation}

In order to obtain an upper bound for $(Mu,Lu)$ in terms of the
$L^2$-norm of $Lu,$ we estimate

\[
(Mu,Lu)\leq \int\int_{\Omega}
\left\{|u|+N\left[|K||u_x|+4|y||u_y|\right]\right\}|Lu|\,dxdy
\]
\[
=\int\int_{\Omega}|u||Lu|\,dxdy+N\int\int_{\Omega}\left[|K||u_x|+4|y||u_y|\right]|Lu|\,dxdy
\]
\begin{equation}\label{new-up-est}
    \equiv i_1+i_2.
\end{equation}
As in case 1, the Schwarz inequality and (\ref{poin}) imply that

\[
i_1 \leq ||u||_{L^2(\Omega)}||Lu||_{L^2(\Omega)} \leq
C||u||_{H^1_0(\Omega;K)}||Lu||_{L^2(\Omega)}.
\]
Because $\Omega$ is bounded, we can fit it inside a rectangle of
the form

\begin{equation}\label{region}
    R= \left\{(x,y)|\gamma_R \leq x \leq \delta_R, \beta_R \leq y\leq
\alpha_R\right\}.
\end{equation}
for sufficiently large values of
$|\alpha_R|,|\beta_R|,|\gamma_R|,$ and $|\delta_R|.$ Define

\[
    T = \max\left\{|\delta_R|+\alpha_R^2,1\right\}.
\]
Then

\[
i_2\leq
N\int\int_{\Omega}\left[\sqrt{T}\sqrt{|K|}|u_x|+4|y||u_y|\right]|Lu|\,dxdy
\]
\[
\leq N\left[\int\int_{\Omega} \Upsilon
\,dxdy\right]^{1/2}||Lu||_{L^2(\Omega)},
\]
where
\[
\Upsilon=T|K||u_x|^2+8\sqrt{T}|\alpha_R|\sqrt{|K|}|u_x||u_y|+16\alpha_R^2|u_y|^2.
\]
Applying Cauchy's inequality to $\Upsilon,$ we obtain
\[
i_2 \leq C||u||_{H^1_0(\Omega;K)}||Lu||_{L^2(\Omega)},
\]
where $C$ depends on $N$ and $R.$

Substituting the estimates for $i_1$ and $i_2$ into inequality
(\ref{new-up-est}), we obtain the desired upper bound for
$(Mu,Lu).$ Combining this with the lower bound (\ref{new-low-est})
and dividing through by the $H^1_0(\Omega;K)$-norm of $u,$ we
complete the proof of Theorem 1.

\bigskip

\textbf{Corollary 2}. \emph{The Dirichlet problem (\ref{f-alt1}),
(\ref{f-alt2}), (\ref{boundary}) with $\kappa\in[0,2]$ possesses a
distribution solution $u\in L^2(\Omega)$ for every $f \in
H^{-1}(\Omega;K).$}

\bigskip

\emph{Proof}. The proof for the case $\kappa=1$ is identical to
the proof of Theorem 2.2 of \cite{LMP}, with Lemma 2.1 of
\cite{LMP} replaced by Theorem 1 of the present communication.
Briefly, we define for $\xi \in C_0^\infty$ a linear functional

\[
J_f(L\xi)=\langle f, \xi \rangle.
\]
This functional is bounded on a subspace of $L^2$ by the
inequality

\begin{equation}\label{schwartz}
    \left |\langle f, \xi \rangle\right| \leq
    ||f||_{H^{-1}\left(\Omega;K\right)}||\xi||_{H_0^1\left(\Omega;K\right)}
\end{equation}
and by Theorem 1 (applied to the second term on the right). Now
standard Hahn-Banach arguments extend the functional to one
defined on all of $L^2.$ The Riesz Representation Theorem then
guarantees the existence of the distribution solution.

If $\kappa\ne1,$ then $L$ is not self-adjoint. Because the
operator adjoint to $L$ has the form

\begin{equation}\label{f-adj}
    L^\ast=\left(x-y^2\right)\frac {\partial^2} {\partial x^2}+\frac{\partial^2}{\partial y^2}+ \left(2-\kappa\right)\frac{\partial}{\partial
    x},
\end{equation}
estimating $L$ for $\kappa$ in $[0,2]$ will also yield estimates
for $L^\ast.$ Applying the preceding argument to the adjoint
operator completes the proof of Corollary 2.

\section{Inequalities leading to weak solutions}

Applications to plasma physics of mixed boundary-value problems
are discussed in \cite{LF}. But we adopt mixed boundary conditions
in this section mostly in the interest of mathematical generality.
The main interest for these equations is in closed Dirichlet and
Neumann problems. The arguments of this section will extend the
results of \cite{O}, for an open weak Dirichlet problem for eq.\
(\ref{cold}), to a class of generalized, closed Neumann problems.

If $\mathbf{u}=\left(u_1,u_2\right)$ and
$\mathbf{w}=\left(w_1,w_2\right)$ are measurable vector-valued
functions on $\Omega,$ then the inner product for the space
$L^2\left(\Omega;\mathbb{R}^2\right)$ will be written

\[
\left(\mathbf{u},\mathbf{w}\right)_{L^2\left(\Omega;\mathbb{R}^2\right)}=\int\int_\Omega\left(u_1w_1+u_2w_2\right)dxdy.
\]
Analogously, we define $\mathcal{H_K}$ to be the Hilbert space of
measurable functions on $\Omega$ for which the weighted $L^2$-norm

\[
||\mathbf{u}||_{H_\mathcal{K}}=\left[\int\int_\Omega\left(|K|u_1^2+u_2^2\right)dxdy\right]^{1/2}
\]
is finite; this norm is induced by the the weighted $L^2$ inner
product

\[
\left(\mathbf{u},\mathbf{w}\right)_\mathcal{K}=\int\int_\Omega\left(|K|u_1w_1+u_2w_2\right)dxdy.
\]
In the notation for these spaces, $\mathcal{K}$ denotes a diagonal
matrix having entries $|K|$ and 1.

By a \emph{weak solution} of a mixed boundary-value problem in this
context we mean an element $\mathbf{u} \in \mathcal{H_K}(\Omega)$
such that

\begin{equation}\label{weak1}
    -\left(\mathbf{u},\mathcal{L}^\ast\mathbf{w}\right)_{L^2\left(\Omega;\mathbb{R}^2\right)}=\left(\mathbf{f},\mathbf{w}\right)_{L^2\left(\Omega;\mathbb{R}^2\right)}
\end{equation}
for every function $\mathbf{w}\in
C^1\left(\overline{\Omega};\mathbb{R}^2\right)$ for which
$\mathcal{K}^{-1}\mathcal{L}^\ast\mathbf{w}\in
L^2\left(\Omega;\mathbb{R}^2\right)$ and for which

\begin{equation}\label{weak2}
    w_1=0\,\forall \left(x,y\right)\in G
\end{equation}
and

\begin{equation}\label{weak3}
    w_2=0\,\forall \left(x,y\right)\in \partial\Omega\backslash G,
\end{equation}
where $G$ is a (possibly empty) subset of $\partial\Omega.$ With a
view toward providing a first-order generalization of eq.\
(\ref{formal}), we choose the differential operator $\mathcal{L}$
to have the form

\begin{equation}\label{op}
\left(
\begin{array}{cc}
  K\partial_x & \partial_y \\
  \partial_y & -\partial_x \\
\end{array}%
\right)+\left(%
\begin{array}{cc}
  \kappa & 0 \\
  0 & 0 \\
\end{array}%
\right).
\end{equation}

\bigskip

\textbf{Theorem 3}. \emph{Let $G$ be a subset of $\partial\Omega$
and let $K=x-y^2.$ Define the functions $b\left(x,y\right)=mK+s$
and $c(y)=\mu y-t,$ where $\mu$ is a positive constant,}

\[
m= \left\{
        \begin{array}{cr}
    \left(\mu + \delta\right)/2 & \mbox{in $\Omega^+$} \\

    \left(\mu - \delta\right)/2 & \mbox{in $\Omega^-$}\\
    \end{array}
    \right.
\]
\emph{for a small positive number $\delta,$ and $t$ is a positive
constant such that $\mu y - t < 0 \, \forall y \in \Omega.$ Let
$s$ be a sufficiently large positive constant. In particular,
choose $s$ to be so large that the quantities $mK+s,$ $2cy+s,$ and
$b^2+Kc^2$ are all positive. Let}

\begin{equation}\label{star1}
    bdy-cdx\leq 0
\end{equation}
\emph{on $G$ and}

\begin{equation}\label{star2}
    K\left(bdy-cdx\right) \geq 0
\end{equation}
\emph{on $\partial\Omega\backslash G.$ Then there exists a
positive constant $C$ such that}

\[
\left(\mathbf{\Psi},\mathcal{L^\ast M}\mathbf{\Psi}\right)\geq
C\int\int_\Omega\left(|K|\Psi_1^2+\Psi_2^2\right)dxdy
\]
\emph{for any sufficiently smooth 2-vector $\mathbf{\Psi},$
provided conditions (\ref{weak2}), (\ref{weak3}) are satisfied on
the boundary for $\mathbf{w}=\mathcal{M}\mathbf\Psi,$ where
$\mathcal{L^\ast}$ is given by (\ref{op}) with $\kappa=1$ and}

\[
\mathcal{M} = \left(%
\begin{array}{cc}
  b & c \\
  -Kc & b \\
\end{array}%
\right).
\]

\bigskip

\emph{Proof.} Again the proof closely follows \cite{LMP} (Lemmas
4.2 and 4.3, and Theorem 4.4). After integration by parts and an
application of the Divergence Theorem, we find that

\[
\left(\mathbf{\Psi},\mathcal{L^\ast
M}\right)_{L^2\left(\Omega;\mathbb{R}^2\right)}=\int\int_\Omega
\left(\alpha\Psi_1^2+2\beta\Psi_1\Psi_2+\gamma\Psi_2^2
\right)dxdy+
\]

\begin{equation}\label{path1}
    \int_{\partial\Omega}\left(\frac{1}{2}Kc\Psi_1^2-b\Psi_1\Psi_2-\frac{c}{2}\Psi_2^2\right)dx+\left(\frac{1}{2}Kb\Psi_1^2+Kc\Psi_1\Psi_2-\frac{1}{2}b\Psi_2^2\right)dy.
\end{equation}
Because $b$ is not continuously differentiable on $\Omega,$ it is
again necessary to introduce a cut along the resonance curve
$x=y^2.$ Evaluating the line integral in (\ref{path1}) for our
choices of $b$ and $c,$ using the fact that $c$ is continuous and
that the discontinuous term in $b$ vanishes on the resonance
curve, we find that the boundary integrals sum to zero along the
cut. Applying the boundary conditions, we obtain

\[
\int_{\partial\Omega}\left(\frac{1}{2}Kc\Psi_1^2-b\Psi_1\Psi_2-\frac{c}{2}\Psi_2^2\right)dx+\left(\frac{1}{2}Kb\Psi_1^2+Kc\Psi_1\Psi_2-\frac{1}{2}b\Psi_2^2\right)dy
\]

\[
    =\frac{1}{2}\int_G\frac{-1}{c^2}\Psi_1^2\left(b^2+Kc^2\right)\left(bdy-cdx\right)
\]

\begin{equation}\label{path2}
+\frac{1}{2}\int_{\partial\Omega\backslash
G}\frac{1}{b^2}\Psi_1^2\left(b^2+Kc^2\right)K\left(bdy-cdx\right).
\end{equation}

The hypotheses insure that the line integrals in (\ref{path2}) are
finite and nonnegative. We have

\[
\alpha = \frac{1}{2}\left[Kb_x-K_xb-\left(Kc\right)_y\right]+
b=\left(\delta/2\right)\left|x-y^2\right|+y\left(\mu
y-t\right)+s/2;
\]

\[
\beta = \frac{1}{2}\left(b_y+c\right)=\left\{
        \begin{array}{cr}
    -\left(1/2\right)\left(\delta y+t\right) & \mbox{in $\Omega^+$} \\

    \left(1/2\right)\left(\delta y-t\right) & \mbox{in $\Omega^-$}\\
    \end{array}
    \right.;
\]

\[
\gamma=\frac{1}{2}\left(c_y-b_x\right)=\left\{
        \begin{array}{cr}
    \left(\mu-\delta \right)/4 & \mbox{in $\Omega^+$} \\

    \left(\mu+\delta \right)/4 & \mbox{in $\Omega^-$}\\
    \end{array}
    \right..
\]
Because $\forall\,\varepsilon>0$

\[
2\beta\Psi_1\Psi_2\geq -\frac{\left(|\delta
y|+|t|\right)^2}{\varepsilon}\Psi_1^2-\varepsilon\Psi_2^2,
\]
it follows that

\[
\int\int_\Omega
\left(\alpha\Psi_1^2+2\beta\Psi_1\Psi_2+\gamma\Psi_2^2\right)dxdy\geq
\]

\[
\int\int_\Omega \left\lbrace\left[\frac{\delta}{2}|K|+y\left(\mu
y-t\right)+\frac{s}{2}-\frac{\left(|\delta
y|+|t|\right)^2}{\varepsilon}\right]\Psi_1^2+\left[\frac{\left(\mu-\delta\right)}{4}-\varepsilon\right]\Psi_2^2\right\rbrace
dxdy
\]

\[
\geq C\int\int_\Omega\left(|K|\Psi_1^2+\Psi_2^2\right)dxdy
\]
for $s$ sufficiently large, $\varepsilon$ sufficiently small, and
$\delta<\mu.$ This completes the proof.

\bigskip

Define the space $\mathcal{H}_{\mathcal{K}^{-1}}(\Omega)$ to
consist of measurable functions $\mathbf{f}=\left(f_1,f_2\right)$
on $\Omega$ for which $\mathcal{K}^{-1}\mathbf{f}$ lies in the
space $L^2\left(\Omega;\mathbb{R}^2\right),$ where

\[
\mathcal{K}^{-1}=\left(%
\begin{array}{cc}
  |K|^{-1} & 0 \\
  0 & 1 \\
\end{array}%
\right).
\]
Then $\mathcal{H}_{\mathcal{K}^{-1}}(\Omega)$ is a Hilbert space
having inner product

\begin{equation}\label{inv}
    \left(\mathbf{v},\mathbf{w}\right)_{\mathcal{H}_{\mathcal{K}^{-1}}(\Omega)}=\left(\mathcal{K}^{-2}\mathbf{v},\mathbf{w}\right)_{L^2\left(\Omega;\mathbb{R}^2\right)}.
\end{equation}

\bigskip

\textbf{Corollary 4}. \emph{Under the hypotheses of Theorem 3,
there exists for every $\mathbf{f}$ such that
$\mathcal{K}^{-1}\mathcal{M}^T \mathbf{f}\in L^2(\Omega)$ a weak
solution to the mixed boundary-value problem
(\ref{weak1})-(\ref{weak3}) with $\mathcal{L}$ given by eq.\
(\ref{op}) with $\kappa=0.$ (The superscripted $T$ denotes matrix
transpose.)}

\bigskip

\emph{Proof}. Apply the proof of Theorem 3 in \cite{M2}
(\emph{c.f.} \cite{LMP}, Lemmas 4.2 and 4.3, and \cite{Pi}),
taking into account that in our case, $\mathcal{L}\ne
\mathcal{L}^\ast.$ Use Theorem 3 of this paper to estimate the
formal adjoint of $\mathcal{L},$ which is obtained by taking
$\kappa=1$ in (\ref{op}). Because
$\mathcal{M}\mathbf\Psi=\mathbf{w},$ we have

\[
\int\int_\Omega\left(|K|\Psi_1^2+\Psi_2^2\right)dxdy=||\mathbf\Psi||_{\mathcal{H}_{\mathcal{K}}\left(\Omega\right)}
\]

\[
=||\mathcal{M}^{-1}\mathbf{w}||_{\mathcal{H}_{\mathcal{K}}\left(\Omega\right)}=||\mathcal{KM}^{-1}\mathbf{w}||_{L^2\left(\Omega;\mathbb{R}^2\right)}
\]
for

\[
\mathcal{K}=\left(%
\begin{array}{cc}
  |K| & 0 \\
  0 & 1 \\
\end{array}%
\right)
\]
and

\[
\mathcal{M}^{-1}\mathbf{w}=\frac{1}{b^2+Kc^2}
\left(%
\begin{array}{c}
  bw_1-cw_2 \\
  cKw_1+bw_2 \\
\end{array}%
\right).
\]
These norms are finite because of the hypotheses on $\mathbf{w}$
and on $b^2+Kc^2.$ Using Theorem 3, we obtain

\[
\frac{\delta}{2}||\mathcal{K}\mathcal{M}^{-1}\mathbf{w}||^2_{L^2\left(\Omega;\mathbb{R}^2\right)}\leq
\left(\mathbf{\Psi},\mathcal{L}^\ast\mathcal{M}\mathbf{\Psi}\right)_{L^2\left(\Omega;\mathbb{R}^2\right)}=\left(\mathcal{M}^{-1}\mathbf{w},\mathcal{L}^\ast\mathbf{w}\right)_{L^2\left(\Omega;\mathbb{R}^2\right)}
\]

\[
=\left(\mathcal{K}\mathcal{M}^{-1}\mathbf{w},\mathcal{K}^{-1}\mathcal{L}^\ast\mathbf{w}\right)_{L^2\left(\Omega;\mathbb{R}^2\right)}\leq
||\mathcal{K}\mathcal{M}^{-1}\mathbf{w}||_{L^2\left(\Omega;\mathbb{R}^2\right)}||\mathcal{K}^{-1}\mathcal{L}^\ast\mathbf{w}||_{L^2\left(\Omega;\mathbb{R}^2\right)},
\]
in which the right-hand side is finite by our definition of
$\mathbf{w}.$ Dividing both sides of this inequality by the
quantity
$||\mathcal{K}\mathcal{M}^{-1}\mathbf{w}||_{L^2\left(\Omega;\mathbb{R}^2\right)},$
we find that

\[
||\mathcal{K}\mathcal{M}^{-1}\mathbf{w}||_{L^2\left(\Omega;\mathbb{R}^2\right)}\leq\frac{2}{\delta}||\mathcal{K}^{-1}\mathcal{L}^\ast\mathbf{w}||_{L^2\left(\Omega;\mathbb{R}^2\right)}.
\]

Define the linear functional

\[
J_f\left(\mathcal{L}^\ast\mathbf{w}\right)=\left(\mathbf{f},\mathbf{w}\right)_{L^2\left(\Omega;\mathbb{R}^2\right)}.
\]
Then

\[
|J_f\left(\mathcal{L}^\ast\mathbf{w}\right)|=|\left(\mathcal{K}^{-1}\mathcal{M}\mathbf{f},\mathcal{KM}^{-1}\mathbf{w}\right)|_{L^2\left(\Omega;\mathbb{R}^2\right)}\leq
C||\mathcal{K}^{-1}\mathcal{L}^\ast\mathbf{w}||_{L^2\left(\Omega;\mathbb{R}^2\right)}.
\]
We conclude that $J_f$ is bounded on the subspace of
$\mathcal{H}_{\mathcal{K}^{-1}}(\Omega)$ consisting of elements
having the form $\mathcal{L}^\ast\mathbf{w}.$ Extending the
operator to a bounded linear functional on the entire space, the
Riesz Representation Theorem guarantees the existence of an
element $\mathbf{v}\in\mathcal{H}_{\mathcal{K}^{-1}}(\Omega)$ for
which

\[
\left(\mathbf{f},\mathbf{w}\right)_{L^2\left(\Omega;\mathbb{R}^2\right)}=\left(\mathbf{v},\mathcal{L}^\ast\mathbf{w}\right)_{\mathcal{H}_{\mathcal{K}^{-1}}(\Omega)}.
\]
The proof is completed by taking
$\mathbf{u}=-\mathcal{K}^{-2}\mathbf{v}$ and applying (\ref{inv}).

\bigskip

\textbf{Remark}. If the vector $\mathbf{u}$ is sufficiently
differentiable, then we can replace $u_1$ by $u_x$ and $u_2$ by
$u_y$ for a scalar function $u\left(x,y\right).$ Formally, we then
obtain from Corollary 4 the existence of a weak solution to eq.\
(\ref{cold}), under mixed boundary conditions consisting of
Dirichlet conditions

\begin{equation}\label{bc1}
    u_xdx+u_ydy=0
\end{equation}
on $G$ and co-normal conditions

\begin{equation}\label{bc2}
    Ku_xdy-u_ydx=0
\end{equation}
on the complement of $G.$

\bigskip

However, the existence of a suitable domain $\Omega$ remains to be
demonstrated.

It is apparent that the conditions on $\Omega$ are non-vacuous if
we consider one of the simplest kinds of piecewise smooth domains,
a box with a vertex at the origin. Let $y_0$ be a positive number,
let $x_0=y_0^2,$ and let $\Omega$ be the rectangle formed by the
line segments

\medskip

$I:$ $0\leq x\leq x_0,$ $y=y_0;$

\medskip

$II:$ $x=0,$ $0\leq y\leq y_0;$

\medskip

$III:$ $0\leq x\leq x_0,$ $y=0;$

\medskip

$IV:$ $x=x_0,$ $0\leq y\leq y_0.$

\medskip

On the line segments $I$ and $II,$ both (\ref{star1}) and
(\ref{star2}) are satisfied. On the line segments $III$ and $IV,$
(\ref{star2}) is satisfied. So we can take $G$ to be a nonempty
subset of $I\cup II.$ Or we can take $G$ to be the empty set, in
which case Theorem 3 guarantees a weak solution to the conormal
problem on $\Omega.$ This problem is hard to solve for equations
of Tricomi type; see \cite{Pi} and the comments in Sec.\ 4 of
\cite{LMP}. In the cold plasma case, it is the weak Dirichlet
problem which is hard to solve, as it is not obvious that there is
a domain on which $G$ can be taken to be the entire boundary.

While the restriction of the boundary arcs to vertical and
horizontal lines obviously simplifies the analysis, it is also
clear that the existence argument for the mixed or co-normal
problem extends to domains more general than a rectangle. On the
elliptic region, all that is required in order for (\ref{star2})
to hold with our choices of  $b$ and $c$ is that $dx/ds$ and
$dy/ds$ both be nonnegative, where $ds$ is the element of arc
length on the boundary. On the hyperbolic region, all that is
required for either (\ref{star1}) or (\ref{star2}) to hold is that
$dx/ds$ and $dy/ds$ both be non-positive. Thus a suitable domain
might have the form of a lens about that segment of the sonic
curve $x=y^2$ which lies in the first quadrant of the $xy$-plane.
It is required that the upper boundary of the lens remain in the
hyperbolic region of the equations without violating the condition
$dy/ds\leq 0.$ For example, let the hyperbolic boundary be given
by the curve $y=x^q$ over the interval $x\in\left[0,1\right],$ for
$q\in\left(0,1/2\right);$ let the elliptic boundary be given by
the curve $y=x^r$ over the interval $x\in\left[0,1\right],$ for
$r>1;$ and let $G$ be a subset of the hyperbolic boundary.

\section{A somewhat smoother class of distribution solutions}

It is also possible to strengthen the result of Sec.\ 2 in a
different direction, by placing hypotheses on $\Omega$ which imply
more smoothness on the part of the distribution solution than mere
square-integrability.

Following Sec.\ 2 of \cite{LP2} we consider a one-parameter family
$\psi_\lambda\left(x,y\right)$ of inhomogeneous dilations given by

\[
\psi_\lambda\left(x,y\right)=\left(\lambda^{-\alpha}x,\lambda^{-\beta}y\right),
\]
where $\alpha, \beta,\lambda \in \mathbb{R}^+.$ These determine an
associated family of operators

\[
\Psi_\lambda u = u\circ \psi_\lambda \equiv u_\lambda.
\]
Denote by $D$ the vector field

\begin{equation}\label{vector}
    Du = \left[\frac{d}{d\lambda}u_\lambda\right]_{|\lambda=1}=-\alpha
x
\partial_x-\beta y\partial y.
\end{equation}
An open set $\Omega \subseteq \mathbb{R}^2$ is said to be
\emph{star-shaped} with respect to the flow of $D$ if
$\forall\left(x_0,y_0\right)\in \overline\Omega$ and each
$t\in\left[0,\infty\right]$ we have
$F_t\left(x_0,y_0\right)\subset \overline\Omega,$ where

\[
F_t\left(x_0,y_0\right)=\left(x(t),y(t)\right)=\left(x_0e^{-\alpha
t},y_0e^{-\beta t}\right).
\]

If $k$ is a given function on $\Omega,$ we define the space
$L^2\left(\Omega;|k|\right)$ and its dual, consisting,
respectively, of functions $u$ for which the norms

\[
||u||_{L^2\left(\Omega;|k|\right)}=\left(\int_\Omega|k|u^2dxdy\right)^{1/2}
\]
and

\[
||u||_{L^2\left(\Omega;|k|^{-1}\right)}=\left(\int_\Omega|k|^{-1}u^2dxdy\right)^{1/2}
\]
are finite; see \cite{LMP}, Sec.\ 3, for details.

Denote by $v$ a $C^1$ solution of the Cauchy problem

\begin{equation}\label{cauchy}
    Hv=u \mbox{ in } \Omega
\end{equation}
with $v$ vanishing on $\partial\Omega\backslash
\lbrace0,0\rbrace,$

\begin{equation}\label{limit}
    \lim_{\left(x,y\right)\rightarrow\left(0,0\right)}v\left(x,y\right)=0,
\end{equation}
and

\begin{equation}\label{oper}
    Hv = av + bv_x+cv_y.
\end{equation}
We assume in the sequel that $v$ exists. This assumption is
justified whenever the following conditions are met: $\Omega$ is
star-shaped with respect to the flow of the vector field
$V=-\left(b,c\right);$ $b=mx$ and $c=\mu y;$ $\mu$ is a positive
constant; $m$ is a step function in $x,$ taking only positive
values, with a single jump at the point $x=0;$ $a$ is a negative
constant having sufficiently large magnitude. A proof of the
sufficiency of these assumptions can be found in step 1 of the
proof of Lemma 3.3, \cite{LMP}.

\bigskip

\textbf{Theorem 5}. \emph{Suppose that $x\geq 0$ on $\Omega$ and
that $\Omega$ is star-shaped with respect to the flow of the
vector field $V=-\left(b,c\right)$ for $b=mx$ and $c=\mu y,$ where
$\mu$ is a positive constant and $m$ exceeds $3\mu.$ Then for
every $u \in C^\infty_0\left(\Omega\right)$ there exists a
positive constant $C$ for which}
\begin{equation}\label{berez}
    ||u||_{L^2\left(\Omega;|k|\right)}\leq
C||Lu||_{H^{-1}\left(\Omega;k\right)},
\end{equation}
\emph{where $k(y)=y^2$ and $L$ satisfies (\ref{f-alt2}) with
$\kappa=1.$}

\bigskip

\emph{Proof.} Let $v$ satisfy eqs.\ (\ref{cauchy})-(\ref{oper}) on
$\Omega$ for $a=-M,$ where $M$ is a positive number satisfying

\[
M=\frac{m-3\mu}{2}-\delta
\]
for some sufficiently small positive number $\delta.$ We have the
integral identities

\begin{equation}\label{intiden}
    \left(Iu,Lu\right)\equiv\left(v,Lu\right)=\left(v,LHv\right).
\end{equation}
Our choice of the coefficients $a,$ $b,$ and $c$ are such that
$a_x=a_y=b_y=c_x=0$ and all second derivatives also vanish.
Substitute into Proposition 12 of the Appendix the quantities
$K=x-y^2,$ $\kappa_1=1,$ $\kappa_2=0,$ $b=b(x),$ $c = c(y).$ We
have $\omega=0,$

\[
\alpha =
K\left(\frac{c_y-b_x}{2}-a\right)+\frac{1}{2}b+\frac{1}{2}K_yc
\]

\[
=\left(\frac{m}{2}-\mu-\delta\right)x+\delta y^2,
\]

\[
\beta =0,
\]
and

\[
\gamma = -a - \frac{c_y}{2}+\frac{b_x}{2}
=M-\frac{\mu-m}{2}=m-2\mu-\delta.
\]
The boundary integral of Proposition 12 vanishes by the compact
support of $u.$ We find that if $\delta$ is sufficiently small
relative to $m$ and $\mu,$ then

\[
    \left(v,LHv\right) \geq \delta\int\int_\Omega\left( y^2 v_x^2 +
    v_y^2\right)
     dxdy.
\]

The upper estimate is immediate. One applies inequality
(\ref{schwartz}) to obtain

\[
\left(v,Lu\right)\leq\left\|v\right\|_{H_0^1\left(\Omega;k\right)}\left\|Lu\right\|_{H^{-1}\left(\Omega;k\right)},
\]
from which the desired inequality follows by the continuity of $H$
as a map from $H_0^1\left(\Omega;k\right)$ into
$L^2\left(\Omega;|k|\right).$ This completes the proof of Theorem
5.

\bigskip

\textbf{Corollary 6}. \emph{Let $\Omega$ be star-shaped with
respect to the flow of the vector field $-V=\left(mx, \mu
y\right),$ where $m$ and $\mu$ are defined as in Theorem 5.
Suppose that $x$ is nonnegative on $\Omega$ and that the origin of
coordinates lies on $\partial\Omega.$ Then for every $f \in
L^2\left(\Omega;|k|^{-1}\right)$ there is a distribution solution
$u\in H_0^1\left(\Omega; k\right)$ to the Dirichlet problem
(\ref{formal}), (\ref{boundary}) where $k=y^2$ and $\kappa=1.$}

\bigskip

\emph{Proof}. The proof mirrors the arguments for the existence of
a distribution solution in the proof of Theorem 3.2 in \cite{LMP},
so we will again be brief. Define a linear functional $J_f$ by the
formula

\[
J_f\left(L\xi\right)=\left(f,\xi\right).
\]
Using the fact that $L$ is self-adjoint for $\kappa=1,$ we
estimate

\[
|J_f\left(L\xi\right)|\leq
C|f||_{L^2\left(\Omega;|k|^{-1}\right)}||L\xi||_{H^{-1}\left(\Omega;k\right)}.
\]
Thus $J_f$ is a bounded linear functional on the subspace of
$H^{-1}\left(\Omega;k\right)$ consisting of elements having the
form $L\xi$ with $\xi\in C_0^\infty\left(\Omega\right).$ Extending
$J_f$ to the entire space, the Riesz Representation Theorem
guarantees the existence of an element $u\in
H^1_0\left(\Omega;k\right)$ for which

\[
\left(u,L\xi\right)=\left(f,\xi\right),
\]
where $\xi\in H^1_0\left(\Omega; k\right).$ This completes the
proof.

\bigskip

\textbf{Remarks}. \emph{i)} While the solution guaranteed by
Corollary 2 is only in $L^2,$ the solution guaranteed by Corollary
6 has a derivative in a weighted $L^2$-space. Thus the solution of
Corollary 6 is closer to a conventional weak solution than is the
solution of Corollary 2.

\medskip

\emph{ii)} The estimates used to obtain inequality (\ref{berez})
extend in a formal way to the weight function $x-y^2$ if, in eq.\
(\ref{f-alt2}), we take $\kappa<0$ with $|\kappa|$ sufficiently
large. Choose

\[
a=\frac{2\delta}{1-\kappa}\left(\frac{5}{2}-2\kappa\right)+\delta\kappa,
\]

\[
b=\frac{4\delta}{1-\kappa}x+mK,
\]
and

\[
c=\frac{2\delta}{1-\kappa}y,
\]
for

\[
m= \left\{
        \begin{array}{cr}
    \delta & \mbox{in $\Omega^+$} \\

    -\delta & \mbox{in $\Omega^-$}\\
    \end{array},
    \right.
\]
$K=x-y^2,$ and $\delta>0.$ In that case, $\omega=0;$

\[
\alpha  = \left\{
        \begin{array}{cr}
    \left[\delta\left(1+2|\kappa|\right)\right]K + 4\delta y^2 & \mbox{in $\Omega^+$} \\

    -\delta K + 4\delta y^2 & \mbox{in $\Omega^-$}\\
    \end{array};
    \right.
\]

\[
\beta =\frac{1}{2}\left[\left(1-\kappa\right)c-b_y\right]=\left\{
        \begin{array}{cr}
    2\delta y & \mbox{in $\Omega^+$} \\

    0 & \mbox{in $\Omega^-$}\\
    \end{array};
    \right.
\]
and

\[
\gamma = -a - \frac{c_y}{2}+\frac{b_x}{2} = \delta
\left(|\kappa|-4 \pm \frac{1}{2}\right).
\]
The cross terms satisfy

\[
2\beta\xi\eta\geq -2\delta\left(y^2\xi^2+\eta^2\right),
\]
so

\[
\int\int_\Omega\left(\alpha\xi^2+2\beta\xi\eta +
\gamma\eta^2\right)dxdy \geq
C\int\int_\Omega\left(|K|\xi^2+\eta^2\right)dxdy
\]
provided $|\kappa|$ is sufficiently large. On the boundary between
$\Omega^+$ and $\Omega^-,$ $K=0,$ so along the cut $b$ is a smooth
function of $x$ only. If $\Omega$ is star-shaped with respect to
the vector field $V=-\left(b,c\right),$ then

\[
    \left(b,c\right)\cdot \widehat{\textbf{n}}= 0,
\]
where $\widehat{\textbf{n}}=\left(-dy,dx\right)$ is the unit
normal to $\partial\Omega^{\pm}.$ Thus we can obtain the integral
identity of Proposition 12 in this case as well. It is not obvious
that the extension is more than formal, as it is not obvious that
the vector field $V$ produces a smooth solution of the system
(\ref{cauchy})-(\ref{oper}).

\subsection{An equation of Keldysh type}

As we remarked in Sec.\ 1, if we allow the resonance curve to be
tangent to a flux surface along an entire interval rather than at
an isolated point, eq.\ (\ref{cold}) can be replaced by an
equation of the form (\ref{cibrario}). We add lower-order terms to
this equation to obtain

\begin{equation}\label{cibrario-alt}
    Lu=x^{2k+1}u_{xx} + u_{yy}+c_1x^{2k} u_x +c_2u=0,
\end{equation}
where we take $k\in\mathbb{Z}^+$ and require that the constants
$c_1$ and $c_2$ satisfy $c_1<k+1$ and $c_2<0$ with $|c_2|$
sufficiently large. Hypotheses on the magnitudes of lower-order
terms are quite common for elliptic-hyperbolic equations, such as
(\ref{cibrario-alt}), which are not of real principle type; see
the remarks in \cite{P1} and compare the conditions on the
lower-order terms in Theorem 4 of \cite{G}.

\bigskip

\textbf{Theorem 7}. \emph{Let a portion of the line $x=0$ lie in
$\Omega$ and let the point $\left(0,0\right)$ lie on
$\partial\Omega.$ Assume that $\Omega$ is star-shaped with respect
to the vector field $V=-\left(b,c\right),$ where $b=mx,$ and
$c=\mu y.$ Let $\mu$ be a positive constant and let}

\[
m= \left\{
        \begin{array}{cr}
    -a/\ell+\mu/2\ell-\delta/\ell & \mbox{in $\Omega^+$} \\

    -a/\ell+\mu/2\ell+\delta/\ell & \mbox{in $\Omega^-$}\\
    \end{array}
    \right.
\]
\emph{for a positive constant $\delta,$ where $\ell= k+1-c_1.$ Let
$a$ be a negative constant of sufficiently large magnitude. In
particular, let $a$ have sufficiently large magnitude that $m$ is
positive. Then for every $w \in C^\infty_0\left(\Omega\right)$
there exists a positive constant $C$ for which}
\[
||w||_{L^2\left(\Omega;|K|\right)}\leq C||L^\ast
w||_{H^{-1}\left(\Omega;K\right)},
\]
\emph{where $K=x^{2k+1}$ and}

\[
    L^\ast w=x^{2k+1}w_{xx}+w_{yy}+
\]

\begin{equation}\label{k1}
    \left(4k+2-c_1\right)x^{2k}w_x+\left[2k\left(2k+1-c_1\right)x^{2k-1}+c_2\right]w
\end{equation}
\emph{is the formal adjoint of the differential operator $L$ of
eq.\ (\ref{cibrario-alt}) acting on $w.$}

\bigskip

\emph{Proof}. The proof is only different in its details from that
of Theorem 5. In Proposition 12 of the Appendix, take
$K=x^{2k+1},$ $\kappa_1 = \left(4k+2-c_1\right)x^{2k}$ and
$\kappa_2 = 2k\left(2k+1-c_1\right)x^{2k-1}+c_2.$ Initially we
perform all operations over $\Omega^+$ and $\Omega^-$
individually. On the interior of these sub-domains the
coefficients are all smooth. We find that

\[
\omega = Mx^{2k-1}+c_2\left(a-m/2-\mu/2\right),
\]
where $M$ is a constant that depends on $k$, $a,$ $m,$ and $c_1$
but not on $c_2.$ Because $a$ is negative and both $m$ and $\mu$
are positive, $\omega$ is positive provided $c_2$ is a negative
number having sufficiently large magnitude relative to the
quantity $M/\left| a-m/2-\mu/2 \right|.$ In addition, Proposition
12 implies that

\[
\alpha =
-K\left(a+2b_x\right)+\frac{3}{2}\left(K_xb+Kb_x\right)+\frac{1}{2}Kc_y-\kappa_1
b
\]

\[
=x^{2k+1}\left[\frac{\mu}{2}-a-\left(k+1-c_1\right)m\right]=\delta|x|x^{2k};
\]

\[
\beta
=\frac{1}{2}\left(K_x-\kappa_1\right)c=\frac{1}{2}\left(c_1-2k-1\right)\mu
yx^{2k};
\]
and

\[
\gamma = -a - \frac{c_y}{2}+\frac{b_x}{2} =
\]

\[
\left\{
        \begin{array}{cr}
    -\left(1+1/2\ell\right)a +\left(1/4\ell-1/2\right)\mu-\delta/2\ell & \mbox{in $\Omega^+$} \\

    -\left(1+1/2\ell\right)a +\left(1/4\ell-1/2\right)\mu+\delta/2\ell & \mbox{in $\Omega^-$}\\
    \end{array}.
    \right.
\]
Then $\gamma\geq \delta$ by the hypotheses on the sign and
magnitude of $a.$ We can write $\beta$ in the form

\[
2\beta = \left(c_1-2k-1\right)\mu y x^{k+1/2}x^{k-1/2}.
\]
Then $\forall\,\varepsilon>0,$

\[
2\beta\xi\eta\geq-\left|c_1-2k-1\right|\mu|y|\left(\varepsilon|x|x^{2k}\xi^2+\frac{1}{\varepsilon}|x|^{2k-1}\eta^2\right).
\]
Choose $\varepsilon$ so small that

\[
\left|c_1-2k-1\right|\mu
\max_{\overline{\Omega}}|y|\varepsilon<\delta
\]
and $|a|$ so large that

\[
\frac{1}{\varepsilon}\max_{\overline{\Omega}}|x|^{2k-1}<-\left(1+\frac{1}{2\ell}\right)a+\left(\frac{1}{4\ell}-\frac{1}{2}\right)\mu-\frac{\delta}{2\ell}.
\]
We can do this, as $|\Omega|$ is bounded and $k\geq 1.$  Then,
arguing as in the proofs of Theorems 1 and 3, we find that

\[
\int\int_\Omega
\left(\alpha\xi^2+2\beta\xi\eta+\gamma\eta^2\right)dxdy\geq
C\int\int_\Omega\left(|K|\xi^2+\eta^2\right)dxdy.
\]
The function $b(x)$ fails to be differentiable on the boundary
between $\Omega^+$ and $\Omega^-.$ But the coefficients of the
boundary terms involving $b_x$ either vanish on the line $K=0$ or
cancel out, and the remaining boundary terms are smooth. Thus we
can integrate over $\Omega^+$ and $\Omega^-,$ using the support of
$v$ and the Divergence Theorem, and obtain

\[
\left(v,LHv\right) \geq C\int\int_\Omega \left(|K| v_x^2 +
v_y^2\right) dxdy.
\]

The remainder of the proof is the same as that of Theorem 5. In
particular, the existence of a suitable function $v$ still follows
from Lemma 3.3 of \cite{LMP}: the roles of $x$ and $y$ in that
argument are symmetric in the sense that the proof is not affected
if the gap in differentiability is shifted from $c(y)$ to $b(x).$

\bigskip

\textbf{Corollary 8}. \emph{Let $\Omega$ and $V$ be defined as in
Theorem 7. Then for every $f \in L^2\left(\Omega;|K|^{-1}\right)$
there is a distribution solution $u$ to the Dirichlet problem
(\ref{cibrario-alt}), (\ref{boundary}) lying in
$H_0^1\left(\Omega;K\right)$ for $K=x^{2k+1}.$}

\bigskip

\emph{Proof}. The proof is obtained by arguing exactly as in the
proof of Corollary 6, substituting the different weight class.

\bigskip

\textbf{Remark}. The weight class of Theorem 7 is more natural
than that of Theorem 5, as we expect any singularities that occur
to be localized on the sonic curve $K.$ However, we do not obtain
uniqueness in any obvious way from either theorem, because we
cannot express weak solutions to either eq.\ (\ref{formal}) or
eq.\ (\ref{cibrario-alt}) as the limit of an appropriately defined
sequence of approximations, as in Sec.\ 3 of \cite{LMP}. We
discuss this problem further in Sec.\ 5.1.

\section{Inequalities leading to unique solutions}

In the case of equations of Tricomi type, one can say much more.
In fact, Lupo, Morawetz, and Payne are able to show the existence,
with a certain degree of regularity, of unique weak solutions to
the Dirichlet problem for equations of Tricomi type (\cite{LMP},
Secs.\ 3 and 5). One problem with adapting that approach to the
present case is the geometry of the sonic curve. The simplest
approach is to introduce a coordinate transformation which
straightens out the sonic curve. One then obtains a problem of
Tricomi type, but the coordinate transformation is singular at the
origin, which is precisely the point of interest in the problem.
In fact, research suggests that the singularity at the origin is
an intrinsic property of the model \cite{PF}, \cite{MSW}, which is
not the case, for example, for the Tricomi equation. This leads us
to expect that solutions to boundary-value problems for eq.\
(\ref{cold}) will live in a rougher space than solutions to
corresponding problems for eq.\ (\ref{tricomi}). In this section
we discuss obstructions to obtaining a theorem on the unique
existence of solutions to closed boundary-value problems for
equations of Keldysh type. We then impose rather harsh conditions
in order to obtain the existence of a unique solution to a
Dirichlet problem for the cold plasma model.

\subsection{On the difficulty of obtaining uniqueness}

A simple but generic example will illustrate the difficulty of
extending, to equations of Keldysh type, methods developed for
proving the weak existence of a unique solution to closed
boundary-value problems for equations of Tricomi type. In our
example the energy estimates used to obtain $H^1_0\left(\Omega;
K\right)$-existence fail for a self-adjoint operator, whereas the
convergence arguments used to obtain uniqueness seem to require a
self-adjoint operator. Obviously, our example says nothing about
whether a unique solution exists, even for this example, but only
about the failure of a direct application of methods developed for
equations of Tricomi type.

Consider an equation of the form

\begin{equation}\label{sa-keldysh}
    \left[K(x)u_x\right]_x+u_{yy}=0,
\end{equation}
where the smooth function $K(x)$ changes type on the line $x=0;$
purely notational alterations $-$ \emph{e.g.}, replacing $x$ by
$\left(x-x_0\right)$ in the definition of $b(x)$ $-$ will extend
the argument to change of type on any vertical ``flux line"
$x=x_0$.

In order to guarantee the existence of a smooth solution to the
system (\ref{cauchy})-(\ref{oper}), we choose $b=mx$ and $c=\mu y$
in that system, where $\mu$ is positive constant and $m$ is a step
function which jumps at the line $x=0$ and has the same sign as
$\mu.$ We take $a$ to be a constant of sufficiently large
magnitude having the sign opposite to that of $\mu$ and $m.$ The
difficulties will not arise from the choices of $\mu$ and $a,$ as
the argument would be the same for any constants. Moreover, the
same difficulties that appear for our choice of vector field will
appear if $b=mx$ is replaced by $b=mf(x),$ where $f$ is any
analytic function vanishing on the flux line (on which $m$ has a
jump discontinuity).

The definition of weak solution introduced in \cite{LMP} for
closed elliptic-hyperbolic boundary-value problems is intermediate
between the distribution solutions studied in the preceding
sections and the strong solutions which we will study in Sec.\
5.2. In accordance with that definition, a \emph{weak solution} to
the Dirichlet problem (\ref{sa-keldysh}), (\ref{boundary}) will be
a function $u \in H^1_0\left(\Omega;K\right)$ for which

\[
\langle Lu, \xi\rangle = \langle f, \xi\rangle
\]
for every $\xi\in H^1_0\left(\Omega;K\right),$ where $L$ is the
differential operator of (\ref{sa-keldysh}). Integrating by parts,

\[
\langle Lu, \xi\rangle\equiv \int\int_\Omega
\left(\left[K(x)u_{x}\right]_x + u_{yy}\right)\xi dxdy =
-\int\int_\Omega\left(Ku_x\xi_x+u_y\xi_y\right)dxdy.
\]
In this case the existence of a weak solution is equivalent to the
existence of a sequence $u_n\in C_0^\infty(\Omega)$ such that

\[
||u_n-u||_{H_0^1\left(\Omega;K\right)}\rightarrow 0 \mbox{ and }
||Lu_n-f||_{H^{-1}\left(\Omega;K\right)}\rightarrow 0
\]
as $n$ tends to infinity. We can thus obtain uniqueness from weak
existence by assuming the existence of two solutions and
subtracting their approximating sequences.

Arguing as in the proof of Theorem 7 in order to establish the
existence of a solution, we estimate the coefficients $\omega,$
$\beta,$ $\alpha,$ and $\gamma.$ We find that $\omega=\beta=0,$
and that

\[
\alpha =
K(x)\left\lbrace-a+\frac{\mu}{2}+\left[\frac{xK'(x)}{K(x)}-1\right]\frac{m}{2}\right\rbrace.
\]
Notice that

\begin{equation}\label{prob}
    \lim_{x\rightarrow 0}\frac{xK'(x)}{K(x)}=\lim_{x\rightarrow
    0}\frac{xK'(x)}{\left[K(x)-K(0)\right]}=1,
\end{equation}
as $K(0)=0.$

In order for eq.\ (\ref{sa-keldysh}) to change type, $K(x)$ must
be monotonic in at least a small interval about $x=0.$ Initially,
suppose that $K'(x)$ is positive near $x=0.$ The sum $-a+\mu/2$
cannot be zero, as $a$ and $\mu$ have been given opposite sign in
order to insure the existence of a solution to
(\ref{cauchy})-(\ref{oper}). If $-a+\mu/2$ is positive, then for
small negative values of $x,$ $K(x)$ will be negative. The
contribution of $\left[\left(xK'/K\right)-1\right]m$ will be small
by (\ref{prob}), so $\alpha$ will be negative. If $-a+\mu/2$ is
negative, then $\alpha$ will be negative for small positive values
of $x$ for the same reason. An analogous argument pertains to the
case in which $K'(x)$ is negative near $x=0.$

This example suggests that proving the uniqueness of weak
solutions to equations of Keldysh type $-$ and to more complicated
non-Tricomi equations $-$ along the lines of \cite{LMP} will
require, at the very least, a quite different choice of vector
field $(b,c),$ which will significantly affect the method.

For these reasons we will derive uniqueness from the existence of
strong solutions rather than from the linearity of the
differential operator on weak solutions. In order to do this, we
change the boundary conditions from closed conditions to open
conditions.

\subsection{Strong solutions}

In the sequel we consider a generalization of the cold plasma
model:

\begin{equation}\label{sys1}
L\mathbf{u}=\mathbf{f}
\end{equation}
for an unknown vector

\[
\mathbf{u}=\left( u_{1}\left( x,y\right) ,u_{2}\left( x,y\right)
\right)
\]
and a given vector

\[
\mathbf{f}=\left( f_{1}\left( x,y\right) ,f_{2}\left( x,y\right)
\right),
\]
where $\left( x,y\right) \in \Omega \subset \mathbb{R}^2.$ Here

\begin{equation}\label{sys2}
    \left( L\mathbf{u}\right) _{1}=\left[ x-\sigma(y) \right] u_{1x} + \kappa_1 u_{1}+\kappa_2 u_{2y},
\end{equation}

\begin{equation}\label{sys3}
    \left( L\mathbf{u}\right) _{2}=u_{1y}-u_{2x},
\end{equation}
where $\kappa_1$ and $\kappa_2$ are constants; $\sigma(y) \geq 0$
is a continuously differentiable function of its argument
satisfying

\begin{equation}\label{sig1}
    \sigma(0)=\sigma'(0)=0,
\end{equation}

\begin{equation}\label{sig2}
    \sigma'(y)>0\,\forall y > 0,
\end{equation}
and
\begin{equation}\label{sig3}
    \sigma'(y)<0\,\forall y < 0.
\end{equation}
In the special case in which $\sigma(y)=y^2,$ $\kappa_2=0,$
$\left(f_1,f_2\right)=\left(f,0\right),$ the components of the
vector $\mathbf{u}$ are continuously differentiable, and
$u_1=u_x,$ $u_2=u_y$ for some twice-differentiable function
$u(x,y),$ the first-order system (\ref{sys1})-(\ref{sys3}) reduces
to eq.\ (\ref{formal}).

We say that a vector $\mathbf{u}=(u_1,u_2)\in L^2(\Omega)$ is a
\textit{strong solution} of an operator equation of the form
(\ref{sys1}), with given boundary conditions, if there exists a
sequence $\mathbf{u}^{\nu }$ of continuously differentiable
vectors, satisfying the boundary conditions, for which
$\mathbf{u}^{\nu }$ converges to $\mathbf{u}$ in $L^2$ and
$L\mathbf{u}^{\nu }$ converges to $\mathbf{f}$ in $L^2.$

A sufficient condition for a vector to be a strong solution was
formulated by Friedrichs \cite{F} (see also \cite{LaP}). An
operator $L$ associated to an equation of the form

\begin{equation}\label{matrixeq}
L\mathbf{u}=A^1\mathbf{u}_x+A^2\mathbf{u}_y+B\mathbf{u},
\end{equation}
where $A^1$, $A^2$, and $B$ are matrices, is said to be
\emph{symmetric-positive} if the matrices $A^1$ and $A^2$ are
symmetric and the matrix

\[
    Q \equiv 2B^* - A_x^1 - A_y^2
\]
is positive-definite, where $B^*$ is the symmetrization of the
matrix $B:$

\[
B^* = \frac{1}{2} \left(B+B^T\right).
\]
If $L$ is not symmetric-positive, then we may consider the
equation

\begin{equation}\label{EL}
    EL\textbf{u}=E\textbf{f}
\end{equation}
for a non-singular matrix $E$ chosen so that $EL$ is
symmetric-positive.

Define the matrix

\begin{equation}\label{bta}
    \beta = n_1A_{|\partial \Omega}^1 + n_2A_{|\partial \Omega}^2,
\end{equation}
where $\mathbf{n}=\left(n_1,n_2\right)$ is the outward-pointing
normal vector on $\partial \Omega.$ Let $\mathcal{N}(x,y),$
$(x,y)\in\partial\Omega,$ be a linear subspace of the vector space
$V,$ where $\textbf{u}:\Omega \cup
\partial \Omega \rightarrow V.$ Suppose that $\mathcal{N}(x,y)$ depends
smoothly on $x$ and $y.$ The boundary condition that $u$ lie in
$\mathcal{N}$ is \emph{admissible} if $\mathcal{N}$ is a maximal
subspace of $V$ and if the quadratic form $(\mathbf{u},\beta
\mathbf{u})$ is non-negative on $\partial \Omega.$

It is sufficient for admissibility that there exist a
decomposition
\[
\beta = \beta_++\beta_-,
\]
for which the following three conditions hold:

\medskip

\emph{i)} The direct sum of the null spaces for $\beta_+$ and
$\beta_-$ spans the restriction of $V$ to the boundary;

\medskip

\emph{ii)} the intersection of the ranges of $\beta_+$ and
$\beta_-$ have only the vector $\mathbf{u}=0$ in common;

\medskip

\emph{iii)} the matrix $\mu=\beta_+-\beta_-$ satisfies
\[
\mu^\ast = \frac{\mu + \mu^T}{2} \geq 0.
\]

\medskip

If these conditions are satisfied, then the boundary condition
\[
\beta_-\textbf{u}=0 \, \mbox{on $\partial \Omega$}
\]
is admissible for eq.\ (\ref{sys1}) and the boundary condition
\[
\mathbf{w}^T \beta_+^T =0 \, \mbox{on $\partial \Omega$}
\]
is admissible for the adjoint problem
\[
L^\ast \mathbf{w}=\mathbf{g} \, \mbox{in $\Omega.$}
\]
Moreover, both problems can be shown to possess unique, strong
solutions.

Write the system (\ref{sys1})-(\ref{sig3}) in the matrix form

\[
L\textbf{u} =\left(%
\begin{array}{cc}
  x-\sigma(y) & 0 \\
  0 & -1 \\
\end{array}%
\right)\left(%
\begin{array}{c}
  u_1 \\
  u_2 \\
\end{array}%
\right)_x + \left(%
\begin{array}{cc}
  0 & 1 \\
  1 & 0 \\
\end{array}%
\right)\left(%
\begin{array}{c}
  u_1 \\
  u_2 \\
\end{array}%
\right)_y
\]

\begin{equation}\label{L}
 +\left(%
\begin{array}{cc}
  \kappa_1 & \kappa_2 \\
  0 & 0 \\
\end{array}%
\right)\left(%
\begin{array}{c}
  u_1 \\
  u_2 \\
\end{array}%
\right).
\end{equation}
We will show the existence of strong solutions to a subclass of
equations for the operator $L.$

\bigskip

\textbf{Theorem 9}. \emph{Assume that on the elliptic boundary of
$\Omega$}

\begin{equation}\label{starlike}
    bn_1+cn_2 \geq 0,
\end{equation}
\emph{where $n_1$ and $n_2$ are components of the outward-pointing
normal vector at each point of $\partial\Omega,$ and where $b$ and
$c$ satisfy, for $K=x-\sigma(y),$ the inequalities}

\begin{equation}\label{Q1}
    2b\kappa_1-b_xK-b+c_yK-c\sigma'(y)>0\, \mbox{in $\Omega$};
\end{equation}

\[
\left(2b\kappa_1-b_xK-b+c_yK-c\sigma'(y)\right)\left(2c\kappa_2+b_x-c_y\right)
\]

\begin{equation}\label{Q2}
    -\left(b\kappa_2+c\kappa_1-c_xK-c-b_y\right)^2>0\,\mbox{in $\Omega$};
\end{equation}

\begin{equation}\label{Q3}
    2c\kappa_2+b_x-c_y>0\,\mbox{in $\Omega$};
\end{equation}

\begin{equation}\label{bQ1}
    bn_1-cn_2 \leq 0 \, \mbox{on $\left(\partial \Omega\right)^-$};
\end{equation}
\emph{and}

\begin{equation}\label{bQ2}
    cKn_1+bn_2\geq 0\, \mbox{on $\left(\partial \Omega\right)^-$},
\end{equation}
\emph{where $\left(\partial \Omega\right)^-$ is the hyperbolic
boundary. Then equation (\ref{sys1}), with $L$ given by (\ref{L})
and the Dirichlet condition}

\begin{equation}\label{dirichlet}
    -u_1n_2+u_2n_1=0
\end{equation}
\emph{imposed on the elliptic portion of $\partial \Omega,$ has a
unique, strong solution on $\Omega$ for every $\mathbf f\in L^2.$}

\bigskip

\emph{Proof}. Define the matrix

\[
    E=\left(%
\begin{array}{cc}
  b & -cK \\
  c & b \\
\end{array}%
\right).
\]
Then the operator $EL$ is symmetric-positive by conditions
(\ref{Q1})-(\ref{Q3}). In order to show the existence of strong
solutions on $\Omega$ it is convenient to produce a decomposition
of the matrix

\begin{equation}\label{bmatrix}
    \beta=\left(%
\begin{array}{cc}
  K\left(bn_1-cn_2\right) & cKn_1+bn_2 \\
  cKn_1+bn_2 & -\left(bn_1-cn_2\right) \\
\end{array}%
\right).
\end{equation}
On the elliptic boundary, choose

\[
\beta_+=\left(%
\begin{array}{cc}
  Kbn_1 & bn_2 \\
  Kcn_1 & cn_2 \\
\end{array}%
\right)
\]

and

\[
\beta_-=\left(%
\begin{array}{cc}
  -Kcn_2 & Kcn_1 \\
  bn_2 & -bn_1 \\
\end{array}%
\right).
\]
Then $\beta_-u=0$ under the boundary condition (\ref{dirichlet}).
Moreover, the intersection and range of the two matrices satisfy
the conditions for admissibility. We have

\[
\mu^\ast=\left(bn_1+cn_2\right)\left(%
\begin{array}{cc}
  K & 0 \\
  0 & 1 \\
\end{array}%
\right),
\]
so condition (\ref{starlike}) implies that Dirichlet conditions
(\ref{dirichlet}) are admissible on the elliptic part of the
boundary.

On the hyperbolic boundary we choose $\beta=\beta_+$ and choose
$\beta_-$ to be the zero matrix. Then the matrix $\mu^\ast$ is
nonnegative on $\left(\partial\Omega\right)^-$ by assumptions
(\ref{bQ1}) and (\ref{bQ2}). Because the other conditions for
admissibility are satisfied trivially on $\left(\partial
\Omega\right)^-,$ the proof of Theorem 9 is complete.

\bigskip

\textbf{Remarks}. \emph{i)} In order to show that inequalities
(\ref{Q1})-(\ref{bQ2}) are non-vacuous, we let $\sigma(y)=y^2,$
$\kappa_1=\kappa_2=0,$ $b=M+NK/2,$ and $c=Ny,$ where $M$ and $N$
are negative constants, $|M|$ is sufficiently large, and
$\Omega\subset \mathbb{R}\times \mathbb{R}\backslash\mathbb{R}^-.$
Then (\ref{Q1})-(\ref{Q3}) are satisfied. Moreover, inequalities
(\ref{bQ1}), (\ref{bQ2}) will be satisfied in a canonical basis
$\left(n_1,n_2\right)=\left(-dy, dx\right)$ provided $dy/ds$ and
$dx/ds$ are both nonpositive. This suggests that, under the
canonical choice of basis, the hyperbolic boundary in Theorem 9
could be a sufficiently thin lens in the first quadrant, the lower
boundary of the lens lying along the sonic curve. As a
particularly simple example, let the hyperbolic boundary be the
arc of the circle

\[
\left(x-1\right)^2+y^2=1
\]
connecting the points $\left(0,0\right)$ and $\left(1,1\right).$
If $|M|$ is sufficiently large, then (\ref{starlike}) requires
only that $dy/ds$ be bounded below away from zero on the elliptic
boundary.

\medskip

\emph{ii)} Under the same choice of basis, in the special case
$u_1=u_x,$ $u_2=u_y,$ we recover condition (\ref{bc1}) from
condition (\ref{dirichlet}) and condition (\ref{bc2}) from the
adjoint condition $Ku_1n_1+u_2n_2=0.$

\medskip

\emph{iii)} A hidden smoothness assumption is contained in the
choice of the component $f_2$ to be zero in eq.\ (\ref{sys1}), as
that would imply, in the case $u_1=u_x,$ $u_2=u_y,$ the
equivalence of mixed partial derivatives of the solution.
Presumably such a condition would be violated at the origin, at
which point the difference of the mixed partial derivatives might
be a delta function. If the difference were somewhat smoother than
a delta function $-$ that is, an $L^2$ function, then the methods
of this section could be applied.

\medskip

\emph{iv)} There is a geometric analogy for condition
(\ref{starlike}): Consider a domain which is star-shaped with
respect to the flow of a given vector field $D$ satisfying
(\ref{vector}). Then the boundary will be \emph{starlike} with
respect to $D$ in the sense that $\alpha n_1 + \beta n_2 \geq 0,$
or, in terms of the basis used in remarks \emph{i)} and
\emph{ii),} $\beta dx-\alpha dy \geq 0,$ on the boundary
(\emph{c.f.} \cite{LP2}). We have avoided imposing the hypothesis
that $\Omega$ is $D$-star-shaped in Theorems 3 and 9, although it
would have been possible to do so formally. The reason is that
equations of the form (\ref{formal}) are only interesting if the
origin is included in the domain, whereas condition $b^2+Kc^2>0$
of Theorem 3 and (\ref{Q1}) of Theorem 9 are problematic if $b$
and $c$ are homogeneous functions passing through the origin.

\section{Weak solutions in $L^2$}

The fact that solutions to the closed boundary-value problems in
Secs.\ 3 and 4 lie in spaces in which a weight function vanishes
at the origin is a strong restriction on their generality. Theorem
9, with the examples given in the remarks following it,
demonstrates the existence of strong solutions which lie in $L^2,$
even at the origin; but unfortunately, the boundary conditions in
that theorem are open. In this section we show that the existence
in $L^2$ of weak solutions to open boundary-value problems is easy
to obtain for a wide class of boundaries by arguments which are
similar to those of \cite{O}.

Define $G$ to be a subset of the non-characteristic portion of the
boundary, $\partial \Omega \backslash \Gamma,$ where $\Gamma$
denotes the part of the boundary consisting of characteristic
lines. Denote by $W(\Omega)$ the linear space of continuously
differentiable functions $ \left( w_{1},w_{2}\right) $ on
$\Omega,$ satisfying $w_{1}=0$ on $G$, $w_2=0$ on $\partial \Omega
\backslash \{\Gamma \cup G\},$

\begin{equation}\label{adjoint}
    w_1dx+w_2dy=0 \, \forall (x,y) \in \Gamma,
\end{equation}
and
\[
\left( L^{\ast }\textbf{w}\right) _{1}=\left[ x-\sigma \left(
y\right) \right] w_{1x}+\left( 1-\kappa_1\right) w_{1}+w_{2y},
\]
\[
\left( L^{\ast }\textbf{w}\right) _{2}=w_{1y}-w_{2x}
\]
in $\Omega.$

We define a \textit{weak solution} to eqs.\
(\ref{sys1})-(\ref{sys3}) with $\kappa_2=0,$ under the mixed
boundary conditions

\begin{equation}\label{dirichlet1}
    u_1dx+u_2dy=0 \, \forall (x,y) \in G,
\end{equation}

\begin{equation}\label{adjoint1}
    \left[ x-\sigma \left( y\right) \right]u_1dy-u_2dx=0 \, \forall (x,y) \in
    \partial \Omega \backslash \{\Gamma \cup G\}
\end{equation}
to be any $\textbf{u}\in L^2(\Omega)$ such that $\forall
\textbf{w}\in W(\Omega),$
\[
\left( \textbf{w},\textbf{f}\right) =-\left( L^{\ast
}\textbf{w},\textbf{u}\right)
\]
under the $L^{2}$ inner product $\left( \;,\;\right) $.

\bigskip

\textbf{Theorem 10}. \emph{Let the noncharacteristic boundary of
$\Omega$ satisfy the differential inequality}

\begin{equation}\label{diffin}
    \frac{dy}{dx}\geq \frac{-ty}{m+x}
\end{equation}
\emph{for a sufficiently large positive constant $m$ and a
constant $t$ exceeding 1. Take the curve $G$ to be the elliptic
boundary of $\Omega.$ Let the constant $\kappa_1$ in eq.\
(\ref{sys2}) exceed 1/2 and let $\kappa_2=0.$ Then
$\forall\,\mathbf{w}\in W(\Omega)$ there exists a positive
constant C for which}

\begin{equation}\label{L2est}
    \left\| \textbf{w}\right\| _{L^2(\Omega)}\leq C\left\| L^{\ast
}\textbf{w}\right\|_{L^2(\Omega)}.
\end{equation}

\bigskip

\emph{Proof}. Define the functions $b= - (m + x)$ and $c=-ty.$ We
will place various conditions on $m,$ all of which require that it
be sufficiently large in comparison with other parameters $-$
$\kappa_1,$ $t,$ $|\Omega|,$ $|\sigma|_{\max(\Omega)}$ and
$|\sigma'|_{\max(\Omega)}$ $-$ as well as with certain explicit
combinations of these parameters. By the continuity of $\sigma,$
we can choose $m$ so large that the matrix
\[
M=\left[
\begin{array}{cc}
b & c \\
-Kc & b
\end{array}
\right]
\]
is non-singular on $\Omega,$ where $K = x-\sigma(y).$ We have
\[
(L^{\ast}\textbf{w},M\textbf{w})=\int\int_{\Omega} Q \,dxdy +
\int\int_{\Omega} S \,dxdy,
\]
where
\[
Q=\alpha w_1^2+2\beta w_1w_2+\gamma w_2^2.
\]
In particular,
\[
\alpha =
\frac{1}{2}K\left(c_y-b_x\right)+\left(\frac{1}{2}-\kappa_1\right)b-\frac{1}{2}\sigma'(y)c
\]
\[
=\left(\kappa_1-\frac{t}{2}\right)x+\frac{t-1}{2}\sigma(y)+m\left(\kappa_1-\frac{1}{2}\right)
+ \frac{ty}{2}\sigma'(y).
\]
We can choose $m$ so large that $\alpha$ is bounded below away
from zero on $\Omega.$ Also,
\[
\gamma = -\frac{1}{2}\left(c_y-b_x\right)=\frac{t-1}{2},
\]
\[
\beta = -\frac{1}{2}\left( Kc_x+ b_y+\kappa_1
c\right)=\frac{\kappa_1 ty}{2},
\]
and $\forall\,\varepsilon>0,$

\[
2\beta\omega_1\omega_2\geq -\left|\kappa_1
ty\right|\left(\frac{\omega_1^2}{\varepsilon}+\varepsilon\omega_2^2\right).
\]
Choose $\varepsilon$ to be so small that $\left|\kappa_1
ty\right|\varepsilon < \left(t-1\right)/2$ and $m$ so large that
$m\left[\kappa_1-(1/2)\right]>\left|\kappa_1
ty\right|/\varepsilon.$ Then there is a positive constant $C$ for
which

\[
\int\int_\Omega Q\,dxdy\geq C
\int\int_\Omega\left(\omega_1^2+\omega_2^2\right)\,dxdy.
\]

Applying the Divergence Theorem, we obtain

\[
\int\int_{\Omega} S \,dxdy=
\]

\[
\int\int_{\Omega}
\left[\left(-K\right)\left(tyw_1w_2+\frac{m+x}{2}w_1^2\right)+\frac{m+x}{2}w_2^2\right]_xdxdy-
\]

\[
\int\int_{\Omega}\left[\frac{1}{2}\left(-K\right)tyw_1^2+(m+x)w_1w_2+\frac{ty}{2}w_2^2\right]_ydxdy=
\]

\[
\int_{\partial\Omega}
\left[\left(-K\right)\left(tyw_1w_2+\frac{m+x}{2}w_1^2\right)+\frac{m+x}{2}w_2^2\right]
dy +
\]

\[
\int_{\partial\Omega}\left[\frac{1}{2}\left(-K\right)tyw_1^2+(m+x)w_1w_2+\frac{ty}{2}w_2^2\right]
dx.
\]

It is not excluded that the hyperbolic boundary may include one or
more characteristic lines $\Gamma.$ Repeatedly applying
(\ref{adjoint}) to the terms in $w_1w_2$ on the boundary integral
over $\Gamma,$ we obtain

\[
\frac{1}{2}\int_\Gamma
\left(Kw_1^2+w_2^2\right)\left[tydx-\left(m+x\right)dy\right]=0,
\]
where on the right we have again used (\ref{adjoint}), and also
the characteristic equations

\[
\frac{dx}{dy}= \pm\sqrt{-K};
\]
\emph{c.f.} \cite{O}, (3.24)$-$(3.26).

On $G,$ $w_1=0$ and the boundary integral reduces to

\[
\frac{1}{2}\int_G w_2^2\left[tydx+(m+x)dy\right].
\]
On $\partial\Omega\backslash \lbrace G\cup\Gamma\rbrace,$ $w_2=0$
and the boundary integral reduces to

\[
\frac{1}{2}\int_{\partial\Omega\backslash \lbrace
G\cup\Gamma\rbrace}w_1^2\left(-K\right)\left[t y
dx+(m+x)dy\right].
\]
Both integrals are nonnegative by (\ref{diffin}).

We have shown that

\[
(L^{\ast}\textbf{w},M\textbf{w})\geq C||\textbf{w}||_{L^2}^2
\]
for some positive number $C.$ Because the elements of $M$ are
bounded on $\Omega,$ applying the Schwarz inequality to the inner
product $(L^{\ast}\textbf{w},M\textbf{w})$ yields for all $
\textbf{w}\in W$ and some new constant $C>0$ inequality
(\ref{L2est}).

\bigskip

\textbf{Corollary 11}. \emph{Under the hypotheses of Theorem 10,
for every} $\textbf{f}\in L^2(\Omega)$ \emph{there exists on
$\Omega$ a weak solution to the mixed boundary-value problem
(\ref{sys1})-(\ref{sys3}), (\ref{dirichlet1}), (\ref{adjoint1}).}

\bigskip

\emph{Proof}. Apply the Riesz Representation Theorem as in
\cite{M1} (\emph{c.f.} \cite{O}).

\bigskip

\textbf{Remarks}. \emph{i)} This class of boundaries suggests the
ice-cream cone-shaped \emph{Tricomi domains} (see, \emph{e.g.},
Sec.\ 2 of \cite{LP1}), rotated by $90^\circ$ in the clockwise
direction (so that the ice-cream cone is lying on its side, with
the cone formed by the intersecting characteristic lines in the
second and third quadrants). This rotation is expected, given the
similarity of eq.\ (\ref{cold}) to the Cinquini-Cibrario equation,
in which the sonic curve is rotated $90^\circ$ with respect to the
sonic curve for the Tricomi equation (\ref{tricomi}). In fact, the
sonic curve of (\ref{cold}) is approximated near the origin by the
sonic curve of the Cinquini-Cibrario equation.

\medskip

\emph{ii)} More generally, Theorem 10 and Corollary 11 remain true
for any choice of $M$ for which

\[
\frac{dy}{dx}\geq -\frac{c}{b}
\]
on the characteristic boundary, $\alpha$ and $\gamma$ are bounded
below by a positive constant, and $\alpha\gamma-\beta^2$ is
nonnegative.

\medskip

\emph{iii)} In the corresponding theorem of \cite{O}, the origin
of coordinates was forced to lie on the boundary of the domain.
The question of whether weak solutions to boundary-value problems
for eqs.\ (\ref{sys1})-(\ref{sys3}) can be shown to exist for
cases in which the origin is allowed to be an interior point was
raised in Ch.\ 3 of \cite{Y}. In allowing the origin to lie at
either a boundary point or an interior point, we have shown the
answer to that question to be ``yes."

\section{Appendix. A multiplier identity}

The proofs of Theorems 5 and 7 are based on a fundamental
multiplier identity for the composition of a second-order operator
and a first-order operator. The proof of the identity is
elementary, as it is based on integration by parts followed by an
application of the Divergence Theorem. But due to its importance,
we provide a full derivation (in slightly greater generality than
we need).

Define the operator $L$ on functions $v\in C^3\left(\Omega\right)$
with $v\equiv 0$ on $\partial\Omega,$ by

\[
Lv=K\left(x,y\right)v_{xx}+v_{yy}+\kappa_1v_x+\kappa_2v,
\]
where the type-change function $K$ and the lower-order
coefficients $\kappa_1=\kappa_1(x,y),$
$\kappa_2=\kappa_2\left(x,y\right)$ are all $C^3$ functions.
Define

\[
Hv=av+bv_x+cv_y,
\]
where $a$ is a constant; $c$ is a linear function of $x$ and $y;$
$b$ is linear in $x$ but possibly nonlinear in $y;$ $b$ and $c$
have vanishing mixed partial derivatives.

\bigskip

\textbf{Proposition 12.}

\[
\int\int_\Omega v\cdot LHv\,dxdy =
\frac{1}{2}\oint_{\partial\Omega}
\left(Kv_x^2+v_y^2\right)\left(cdx-bdy\right)
\]

\[
+ \int\int_\Omega\omega v^2+\alpha v_x^2 + 2\beta v_xv_y+ \gamma
v_y^2
     dxdy,
\]
\emph{where}

\[
2\omega =
\left(K_{xx}-\kappa_{1x}+2\kappa_2\right)a-\left[\left(K_{xx}-\kappa_{1x}+\kappa_2\right)b\right]_x
\]

\[
-\left[\left(K_{xx}-\kappa_{1x}+\kappa_2\right)c\right]_y;
\]

\[
\alpha =
\left(\frac{c_y-b_x}{2}-a\right)K+\left(\frac{3}{2}K_x-\kappa_1\right)b+\frac{c}{2}K_y;
\]

\[
2\beta = \left(K_x-\kappa_1\right)c-\left(c_xK+b_y\right);
\]

\[
\gamma = \frac{b_x-c_y}{2}-a.
\]

\bigskip

\emph{Proof.} Writing

\[
LHv=K\left[\left(a+2b_x\right)v_{xx}+bv_{xxx}+2c_xv_{yx}+cv_{yxx}\right]
\]

\[
+\left(a+2c_y\right)v_{yy}+b_{yy}v_x+2b_yv_{xy}+bv_{xyy}+cv_{yyy}
\]

\[
+\kappa_1\left[\left(a+b_x\right)v_x+bv_{xx}+c_xv_y+cv_{yx}\right]+\kappa_2\left(av+bv_x+cv_y\right),
\]
we have

\[
v\cdot LHv=\sum_{i=1}^{16}\tau_i,
\]
where

\[
\tau_1=vK\left(a+2b_x\right)v_{xx}=\left\lbrace\left[\left(a+2b_x\right)\left(Kv_x-\frac{1}{2}K_xv\right)\right]v\right\rbrace_x
\]

\[
-K\left(a+2b_x\right)v_x^2+\frac{1}{2}K_{xx}\left(a+2b_x\right)v^2;
\]

\[
\tau_2 = vKbv_{xxx}=
\]

\[
\left\lbrace
-\frac{1}{2}Kbv_x^2+\left[b\left(Kv_{xx}-K_xv_x+\frac{1}{2}K_{xx}v\right)+b_x\left(K_xv-Kv_x\right)\right]v\right\rbrace_x
\]

\[
-\frac{1}{2}\left(K_{xxx}b+3K_{xx}b_x\right)v^2+\frac{3}{2}\left(Kb\right)_xv_x^2;
\]

\[
\tau_3=2vKc_xv_{yx}=2\left(vKc_xv_y\right)_x-\left(K_xc_xv^2\right)_y-2Kc_xv_xv_y+K_{xy}c_xv^2;
\]

\[
\tau_4=vKcv_{yxx}=
\]

\[
\left\lbrace
v\left[c\left(Kv_{yx}-K_xv_y\right)-Kc_xv_y\right]\right\rbrace_x-
\left\lbrace\frac{1}{2}Kcv_x^2-\left(\frac{1}{2}K_{xx}c+K_xc_x\right)v^2\right\rbrace_y
\]

\[
-\frac{1}{2}\left\lbrace\left[K_{xxy}c+K_{xx}c_y+2K_{xy}c_x\right]v^2-\left(Kc\right)_yv_x^2\right\rbrace+\left(Kc\right)_xv_xv_y;
\]

\[
\tau_5=v\left(a+2c_y\right)v_{yy}=\left[v\left(a+2c_y\right)v_y\right]_y-\left(a+2c_y\right)v_y^2;
\]

\[
\tau_6=vb_{yy}v_x=\frac{1}{2}\left(b_{yy}v^2\right)_x;
\]

\[
\tau_7 = 2vb_yv_{xy}= \left(2vb_yv_x\right)_y-2b_yv_xv_y
-\left(b_{yy}v^2\right)_x;
\]

\[
\tau_8 = vbv_{xyy} = -\left(\frac{1}{2}bv_y^2\right)_x
+\left[\left(bv_{xy}-b_yv_x\right)v\right]_y +
\frac{1}{2}\left[b_xv_y^2+\left(b_{yy}v^2\right)_x\right]+b_yv_xv_y;
\]

\[
\tau_9 = vcv_{yyy}=
-\frac{1}{2}\left(cv_y^2\right)_y+\left[\left(cv_{yy}-c_yv_y\right)v\right]_y+\frac{3}{2}c_yv_y^2;
\]

\[
\tau_{10} =
v\kappa_1\left(a+b_x\right)v_x=\frac{1}{2}\left[\left(a+b_x\right)\kappa_1v^2\right]_x-\frac{1}{2}\kappa_{1x}\left(a+b_x\right)v^2;
\]

\[
\tau_{11} = v\kappa_1bv_{xx} =
\]

\[
\left\lbrace\left[\kappa_1bv_x-\frac{1}{2}\left(\kappa_{1}b\right)_xv\right]
v\right\rbrace_x +
\frac{1}{2}\left(\kappa_{1xx}b+2\kappa_{1x}b_x\right)v^2-\kappa_1bv_x^2;
\]

\[
\tau_{12} = v\kappa_1c_xv_y =
\frac{1}{2}\left[\left(\kappa_1c_xv^2\right)_y-\kappa_{1y}c_x\right]v^2;
\]

\[
\tau_{13} =
v\kappa_1cv_{yx}=\left(v\kappa_1cv_y\right)_x-\frac{1}{2}\left[\left(\kappa_1c\right)_xv^2\right]_y
\]

\[
-\kappa_1cv_xv_y+\frac{1}{2}\left[\kappa_{1y}c_x+\left(\kappa_{1x}c\right)_y\right]v^2;
\]

\[
\tau_{14}=\kappa_2av^2;
\]

\[
\tau_{15}=v\kappa_2bv_x=\frac{1}{2}\left(\kappa_2bv^2\right)_x-\frac{1}{2}\left(\kappa_2b\right)_xv^2;
\]

\[
\tau_{16}=v\kappa_2cv_y=\frac{1}{2}\left(c\kappa_2v^2\right)_y-\frac{1}{2}\left(c\kappa_2\right)_yv^2.
\]

Collect terms and integrate over $\Omega.$ Applying the Divergence
Theorem, taking into account that $v$ (but not necessarily $v_x$
or $v_y$) vanishes on $\partial\Omega,$ completes the proof.

\bigskip

Similar estimates are applied, in the proof of Theorem 1, to a
product having the simpler form $(Hv,Lv).$ The matrix identities
which underlie the proofs of Theorems 3 and 10 are analogous but
also simpler, as the differential operators in those cases are
first-order and the equations are in reduced form. In Proposition
12 the operator is effectively third-order and there are
zeroth-order terms in the equations.

\bigskip

\textbf{Acknowledgment}. I am grateful to Profs.\ D.\ Lupo and K.\
R.\ Payne for discussion of this problem, and to the referee for
extremely helpful comments.

\end{document}